\documentclass[trackchanges]{aastex701}

\usepackage{hyperref}
\usepackage{amsmath}
\usepackage{bm}

\DeclareMathOperator{\arsinh}{arsinh}
\newcommand{\Nobs}{{N_\mathrm{obs}}}

\newcommand{\average}[1]{{\langle #1 \rangle}}
\newcommand{\diag}[1]{{\operatorname{diag}(#1)}}
\newcommand{\expect}[1]{{\mathbb{E}\left[#1\right]}}
\newcommand{\variance}[1]{{\mathbb{V}\left[#1\right]}}
\newcommand{\T}{{\mathrm{T}}}

\shorttitle{Photometric Uncertainty Correction for Time-Domain Surveys}
\shortauthors{Malanchev et al.}

\begin{document}

\title{Photometric Uncertainty Correction for Time-Domain Surveys: Vera C. Rubin Data Preview 2 Case Study}

\author[0000-0001-7179-7406]{Konstantin Malanchev}
\affiliation{The McWilliams Center for Cosmology \& Astrophysics, Department of Physics, Carnegie Mellon University, Pittsburgh, PA 15213, USA}
\email[show]{malanchev@cmu.edu}

\author[0009-0009-7822-7110]{Doug Branton}
\affiliation{DiRAC Institute and the Department of Astronomy, University of Washington, 3910 15th Ave NE, Seattle, WA 98195, USA}
\email{brantd@uw.edu}

\author[0009-0007-9870-9032]{Sandro Campos}
\affiliation{The McWilliams Center for Cosmology \& Astrophysics, Department of Physics, Carnegie Mellon University, Pittsburgh, PA 15213, USA}
\email{scampos@andrew.cmu.edu}

\author[0000-0003-3287-5250]{Neven Caplar}
\affiliation{DiRAC Institute and the Department of Astronomy, University of Washington, 3910 15th Ave NE, Seattle, WA 98195, USA}
\email{ncaplar@uw.edu}

\author[0000-0002-5995-9692]{Mi~Dai}
\affiliation{
        Pittsburgh Particle Physics, Astrophysics, and Cosmology Center (PITT PACC).
        Physics and Astronomy Department, University of Pittsburgh,
        Pittsburgh, PA 15260, USA
}
\email{mi.dai@pitt.edu}

\author[0000-0002-1074-2900]{Melissa DeLucchi}
\affiliation{The McWilliams Center for Cosmology \& Astrophysics, Department of Physics, Carnegie Mellon University, Pittsburgh, PA 15213, USA}
\email{delucchi@andrew.cmu.edu}

\author[0000-0001-5028-146X]{Olivia Lynn}
\affiliation{The McWilliams Center for Cosmology \& Astrophysics, Department of Physics, Carnegie Mellon University, Pittsburgh, PA 15213, USA}
\email{}

\author[0000-0003-2271-1527]{Rachel Mandelbaum}
\affiliation{The McWilliams Center for Cosmology \& Astrophysics, Department of Physics, Carnegie Mellon University, Pittsburgh, PA 15213, USA}
\email{rmandelb@andrew.cmu.edu}

\author[0009-0005-8764-2608]{Sean McGuire}
\affiliation{The McWilliams Center for Cosmology \& Astrophysics, Department of Physics, Carnegie Mellon University, Pittsburgh, PA 15213, USA}
\email{seanmcgu@andrew.cmu.edu}

\author[0000-0001-9547-1347]{Inar Timiryasov}
\affiliation{Niels Bohr Institute, University of Copenhagen, Blegdamsvej 17, 2100 Copenhagen, Denmark}
\email{inar.timiryasov@nbi.ku.dk}

\begin{abstract}
Time-domain science, from variability detection to supernova cosmology, relies on accurate photometric uncertainties.
We present a data-driven, catalog-level method to validate and improve the photometric uncertainties reported by time-domain surveys.
The analysis is based on transforming light curves of non-variable objects into mean-variance whitened signals.
We then estimate a correction to the uncertainties by matching the distribution of the whitened signals of all objects to the standard normal distribution, which they ought to follow when the uncertainties are correct.
The correction can be modeled as a function of each measurement's flux, reported uncertainty, and available observational metadata, and can be applied to measurements of both non-variable and variable objects.
Applied to the Rubin Observatory Data Preview~2 (DP2), the method shows that the uncertainties of bright sources are underestimated by up to a factor of five.
The inferred corrections correspond to a magnitude error floor of about 10\,mmag for both direct-image and difference-image photometry, decreasing to about 6\,mmag in deeper fields.
We provide the corrected uncertainties as part of the Rubin DP2 HATS catalogs.
\end{abstract}


\section{Introduction}\label{sec:intro}

Time-domain photometric surveys study the variable sky by measuring the flux of each object many times, which gives its light curve.
Light curves let us follow how individual stars, galaxies, and transients change in time, discover rare phenomena, and study the variability of entire populations statistically.
The Vera C. Rubin Observatory Legacy Survey of Space and Time \citep[LSST;][]{2019ApJ...873..111I} will obtain light curves of billions of objects over ten years.
Each flux measurement comes with an uncertainty estimated by the survey photometric pipeline \citep[e.g.,][]{2018AJ....155...41B, 2018PASJ...70S...5B, 2019ASPC..523..521B, 2019PASP..131a8003M}.
These uncertainties are derived from a noise model of the image: photon noise of the source and the sky background, detector read noise, and their propagation through the point spread function (PSF) or aperture flux measurement and, for the difference-image photometry, through the image subtraction.
Such noise models do not, and often cannot, include every source of scatter, for example, errors of the photometric calibration, flat-fielding, and PSF models, astrometric errors of the source position, pixel noise correlations introduced by resampling and convolution, contamination by neighboring sources, and image subtraction artifacts \citep{2015AJ....150..172K, 2023arXiv230516279M}.
As a result, the reported uncertainties may be systematically underestimated, and the degree of miscalibration may depend on the source brightness, the observing conditions, and the position of the source on the detector.

Correct uncertainties are essential for many time-domain applications.
Variability detection usually compares the observed scatter of a light curve with its reported uncertainties, for example, via the $\chi^2$ of a constant-flux model \citep{2007AJ....134.2236S, 2017MNRAS.464..274S}, and the excess variance of active galactic nuclei (AGN) light curves is measured as the difference between these two quantities \citep{2003MNRAS.345.1271V}; underestimated uncertainties lead to false variability detections, and overestimated ones hide low-amplitude variability.
The reported uncertainties also define the detection significance of transients and the forced-photometry upper limits \citep{2023arXiv230516279M}, as well as the goodness of fit of light curve models, for instance, of microlensing events \citep{2016AcA....66....1S}.
Type~Ia supernova (SN~Ia) cosmology is particularly sensitive to photometric systematics: the photometric uncertainties propagate into the covariance of the distance moduli \citep{2022ApJ...938..110B, 2024ApJ...973L..14D, 2025ApJ...986..231R}, and the reported uncertainties of the Dark Energy Survey Supernova Program difference-image and scene-modeling photometry required corrections as a function of the host galaxy surface brightness at the transient location \citep{2015AJ....150..172K, 2019ApJ...874..106B, 2024ApJ...975....5S}.

Photometric pipelines are commonly tested with source injection, in which synthetic sources of known flux are added to the real images and processed with the same pipeline \citep[see, e.g.,][]{1987PASP...99..191S, 2015AJ....150..172K, 2018PASJ...70S...6H}.
Comparing the measured fluxes of the injected sources with their true values characterizes the detection efficiency, the photometric biases, and the flux scatter.
Because the injected sources share the noise, the background, and the neighbors of the real ones, such tests reveal, for example, the additional flux scatter of transients on bright host galaxies in the difference images \citep{2015AJ....150..172K} and the larger uncertainties and biased fluxes of blended sources \citep{2018PASJ...70S...6H}.
Since the synthetic sources are rendered with the PSF model and the photometric calibration determined by the pipeline, source injection characterizes the measurement given these models and, by design, is not sensitive to their own uncertainties, such as errors of the photometric zero points or of the PSF model.
It also involves reprocessing a substantial fraction of the images, which makes it computationally demanding for large time-domain surveys, so it is complementary to validation based on the measurements themselves.

When the uncertainties are validated at the catalog level, the natural reference is provided by repeated measurements of non-variable sources, for which any scatter beyond the reported uncertainties indicates their miscalibration.
A common practice is to compare the root-mean-square scatter of light curves with their typical reported uncertainty, as was done for the Spitzer light curves of young stellar objects \citep{2014AJ....148...92R} and the DECam photometry of the High Cadence Transient Survey \citep[HiTS;][]{2018PhDT.......286M}.
Alternatively, each measurement may be normalized by its own reported uncertainty, and the resulting $\chi^2$ compared with its expected distribution.
For the SDSS Stripe~82, this approach showed that the reported uncertainties of the PSF photometry of non-variable stars are consistent with the observed scatter \citep{2007AJ....134..973I}.
Another analysis of the same stripe found the reported uncertainties underestimated by a factor of 1.2 to over 2, depending on the magnitude, with the largest discrepancy for the model and composite-model magnitudes \citep{2005astro.ph..8564S}.
A parametric noise model may also be fitted to the light curve scatter and then applied to each measurement with its own reported uncertainty: for the OGLE-IV photometry, the reported uncertainties are multiplied by a scale factor and an error floor is added in quadrature, with both parameters fitted for every field and detector \citep{2016AcA....66....1S}; the per-band constant magnitude error model we use in this paper (Section~\ref{sec:models}) has a similar form.
The uncertainties of individual light curves are often rescaled by the square root of the reduced $\chi^2$ of the flux in a part of the light curve where the source is expected to be constant, for example, before a transient event; this is recommended for the ZTF forced photometry \citep{2019ApJ...886..152Y, 2023arXiv230516279M}.
These catalog-level approaches reduce each light curve to its scatter or $\chi^2$, which constrains mainly the width of the distribution of the deviations, and calibrate the correction either for each light curve separately or in bins of a few light-curve properties, such as the magnitude and the detector.
Light curves of non-variable objects contain more information: the normalized deviation of every measurement may be kept instead of being summed over the light curve, the distribution of these deviations may be compared with the expected one as a whole, and the correction may depend on many properties of the individual measurements.

In this paper, we develop a method based on the deviations of individual measurements.
To keep every measurement in the analysis, we mean-variance whiten the light curves of non-variable objects: the resulting whitened signal follows the standard normal distribution for correct uncertainties, regardless of the light curve length and of how the uncertainties change from visit to visit.
We compare the whole shape of its distribution with the expected one through their characteristic functions.
The resulting statistic is differentiable, so it serves as a loss function for fitting the uncertainty correction as a function of many properties of individual measurements, such as their flux, reported uncertainty, and observing conditions.
The correction model may be of any complexity, from a simple analytical model with a few parameters, such as a per-band magnitude error floor, to a neural network.
The fitted correction can be applied to new measurements, including those of variable objects.
We do not address correlated systematic errors shared by many light curves \citep[e.g.,][]{2005MNRAS.356.1466T}, and assume that measurements of different objects are independent.

We demonstrate the method on the Rubin Observatory Data Preview~1 \citep[DP1;][]{2026AJ....171..360V} and Data Preview~2 (DP2) data, for both the direct-image and the difference-image forced photometry.
This demonstration is not intended to be a definitive validation of the Rubin photometric pipeline, which is still being actively developed, but it shows how the method performs on real survey data and which simple corrections it suggests.
The paper is organized as follows.
Section~\ref{sec:data} describes the data, Section~\ref{sec:methods} describes the method, Section~\ref{sec:results} presents the results for DP1 and DP2, Section~\ref{sec:discussion} discusses alternative setups, and Section~\ref{sec:conclusions} concludes.
The derivation of the whitening operator is given in Appendix~\ref{sec:light-curve-whitening}.

\section{Data}\label{sec:data}

We use the Vera C. Rubin Observatory Data Preview 1 \citep[DP1;][]{2026AJ....171..360V, 2025lsst.data....3N} and Data Preview 2 \citep[DP2;][]{2026rubn.rept....2V, 2026lsst.data...42N}, which are released as part of the Rubin Early Science Program \citep{2025rubn.rept...79G}.
We access them as Hierarchical Adaptive Tiling Scheme (HATS) catalogs\footnote{A list of HATS catalogs for different surveys is available at \url{https://data.lsdb.io}.} with \texttt{LSDB} \citep{2025arXiv250102103C}, converted from the catalog files provided by the Rubin Observatory.
In these catalogs, the light curves are stored as nested columns of the object table, so each row contains both the object-level properties and the forced source measurements, on the direct and difference images.
Rows without valid coordinates are removed, the mid-exposure time of the visit, \texttt{midpointMjdTai}, is joined from the visit table to each source, and AB magnitudes and their errors are added for each PSF flux and flux error column \citep{2025arXiv250623955M}.
The catalogs are available to Rubin data rights holders via the Rubin Science Platform and the Independent Data Access Centers.

In the analysis, we exclude the forced sources that have any of the following flags set: \texttt{psfFlux\_flag}, the general failure flag of the PSF flux measurement, and \texttt{pixelFlags\_bad}, \texttt{pixelFlags\_saturated}, \texttt{pixelFlags\_cr}, and \texttt{pixelFlags\_suspect}, which mark bad, saturated, cosmic-ray-affected, and suspect pixels, respectively.
For the difference-image photometry, we additionally require \texttt{psfDiffFlux\_flag} to be unset, keeping the \texttt{psfFlux\_flag} cut, because the direct-image flux is one of the model input features (Section~\ref{sec:features}).
We also exclude the light curves of the objects without a PSF magnitude or extendedness in the given band.

\subsection{Rubin DP1}

DP1 is based on 1792 exposures obtained over 48 nights in late 2024 with the LSST Commissioning Camera \citep[LSSTComCam;][]{2024rubn.inst....2S}, and covers about 15\,deg$^2$ in seven fields, each observed in the six bands, $ugrizy$ \citep{2026AJ....171..360V}.
We use the DP1 HATS catalogs\footnote{\url{https://lsdb.io/dp1}} introduced by \citet{2025arXiv250623955M}.

The DP1 object catalog contains 2,299,726 objects and 268,796,943 PSF forced sources across all bands, before any selection.
After the selection described in Section~\ref{sec:rejection-of-variable-objects} (source quality cuts, rejection of variable objects, and at least 10 sources per single-band light curve), DP1 provides 1,408,102 trainable single-band light curves with 42,101,420 sources in total for the calibrated-image (direct) PSF photometry.
Table~\ref{tab:dp1-counts} lists the counts per band and per train/validation/test split; the splits are defined by the hash boundaries of 0.6 and 0.85 (Section~\ref{sec:implementation}).
The same selection applied to the difference-image PSF photometry yields 1,467,450 trainable single-band light curves with 44,713,901 sources; Table~\ref{tab:dp1-counts-diff} lists the corresponding per-band and per-split counts.

\begin{deluxetable}{lr@{\;}rr@{\;}rr@{\;}rr@{\;}r}
\tablecaption{Trainable single-band light curves and sources in DP1 direct-image forced source photometry.\label{tab:dp1-counts}}
\tablehead{\colhead{Band} & \multicolumn{2}{c}{Train} & \multicolumn{2}{c}{Validation} & \multicolumn{2}{c}{Test} & \multicolumn{2}{c}{Total}}
\startdata
$u$ & 57,610 & (1,098,410) & 23,895 & (457,434) & 14,281 & (273,326) & 95,786 & (1,829,170) \\
$g$ & 153,479 & (5,123,021) & 63,960 & (2,143,609) & 38,739 & (1,284,815) & 256,178 & (8,551,445) \\
$r$ & 191,287 & (6,235,470) & 79,658 & (2,606,288) & 47,975 & (1,558,999) & 318,920 & (10,400,757) \\
$i$ & 277,632 & (7,357,290) & 116,858 & (3,092,408) & 70,436 & (1,867,266) & 464,926 & (12,316,964) \\
$z$ & 119,866 & (4,701,015) & 49,674 & (1,950,020) & 29,821 & (1,167,931) & 199,361 & (7,818,966) \\
$y$ & 43,796 & (711,969) & 18,192 & (294,659) & 10,943 & (177,490) & 72,931 & (1,184,118) \\
\hline
Total & 843,670 & (25,227,175) & 352,237 & (10,544,418) & 212,195 & (6,329,827) & 1,408,102 & (42,101,420) \\
\enddata
\tablecomments{Forced PSF photometry on the calibrated science (direct) images. Light curves passing the source quality cuts and the variability rejection of Section~\ref{sec:rejection-of-variable-objects}, with at least 10 sources in the given band. The number of sources is given in parentheses.}
\end{deluxetable}

\begin{deluxetable}{lr@{\;}rr@{\;}rr@{\;}rr@{\;}r}
\tablecaption{Trainable single-band light curves and sources in DP1 difference-image forced source photometry.\label{tab:dp1-counts-diff}}
\tablehead{\colhead{Band} & \multicolumn{2}{c}{Train} & \multicolumn{2}{c}{Validation} & \multicolumn{2}{c}{Test} & \multicolumn{2}{c}{Total}}
\startdata
$u$ & 58,068 & (1,106,567) & 24,019 & (459,564) & 14,299 & (274,079) & 96,386 & (1,840,210) \\
$g$ & 161,170 & (5,500,923) & 67,150 & (2,273,457) & 40,701 & (1,368,090) & 269,021 & (9,142,470) \\
$r$ & 198,812 & (6,582,231) & 82,945 & (2,773,089) & 49,888 & (1,646,405) & 331,645 & (11,001,725) \\
$i$ & 289,625 & (7,781,537) & 121,903 & (3,268,062) & 73,188 & (1,963,759) & 484,716 & (13,013,358) \\
$z$ & 127,558 & (5,121,279) & 53,009 & (2,130,842) & 31,893 & (1,273,949) & 212,460 & (8,526,070) \\
$y$ & 43,976 & (715,650) & 18,261 & (296,087) & 10,985 & (178,331) & 73,222 & (1,190,068) \\
\hline
Total & 879,209 & (26,808,187) & 367,287 & (11,201,101) & 220,954 & (6,704,613) & 1,467,450 & (44,713,901) \\
\enddata
\tablecomments{As Table~\ref{tab:dp1-counts}, but for the forced PSF photometry on the difference images. The number of sources is given in parentheses.}
\end{deluxetable}

\subsection{Rubin DP2}

DP2 \citep{2026rubn.rept....2V} uses the LSST Camera \citep[LSSTCam;][]{2025rubn.inst...36S} data, mostly obtained during the commissioning in 2025 April--September, with a smaller number of later exposures, taken until 2026 January in the same sky regions.
The DP2 HATS catalogs were converted from the Rubin catalog files with the same pipeline as the DP1 ones \citep{2025arXiv250623955M}\footnote{\url{https://github.com/lincc-frameworks/rubin-dash}}.
The DP2 object catalog contains 786,420,720 objects in 8,840 HATS partitions, 3.6\,TiB on disk.
The DP2 DIA object catalog, which contains the objects detected on the difference images by the difference image analysis (DIA) pipeline, has 232,004,216 objects in 8,393 HATS partitions, 1.1\,TiB on disk; its light curves are the forced PSF photometry at the DIA object positions, \texttt{diaObjectForcedSource}, on both the direct and the difference images.
We train the models on the object catalog only and apply them to both catalogs (Section~\ref{sec:data-corrected}).

After the selection of Section~\ref{sec:methods} (source quality cuts, rejection of variable objects, and at least 10 sources per single-band light curve), DP2 provides 435,789,499 trainable single-band light curves with 7,923,228,937 sources in total for the calibrated-image (direct) PSF photometry.
Table~\ref{tab:dp2-counts} lists the counts per band and per train/validation/test split; the splits are defined by the hash boundaries of 0.7 and 0.85 to keep the validation set small enough to be re-evaluated continuously (Section~\ref{sec:implementation}).
The same selection applied to the difference-image PSF photometry yields 457,964,901 trainable single-band light curves with 8,838,079,145 sources; Table~\ref{tab:dp2-counts-diff} lists the corresponding per-band and per-split counts.

\begin{deluxetable}{lr@{\;}rr@{\;}rr@{\;}rr@{\;}r}
\tablecaption{Trainable single-band light curves and sources in DP2 direct-image forced source photometry.\label{tab:dp2-counts}}
\tablehead{\colhead{Band} & \multicolumn{2}{c}{Train} & \multicolumn{2}{c}{Validation} & \multicolumn{2}{c}{Test} & \multicolumn{2}{c}{Total}}
\startdata
$u$ & 3,361,446 & (166,102,460) & 719,542 & (35,539,740) & 720,210 & (35,713,358) & 4,801,198 & (237,355,558) \\
$g$ & 14,340,193 & (472,678,640) & 3,071,989 & (101,352,913) & 3,072,876 & (101,286,493) & 20,485,058 & (675,318,046) \\
$r$ & 63,014,560 & (1,165,018,873) & 13,510,135 & (249,683,127) & 13,499,681 & (249,479,189) & 90,024,376 & (1,664,181,189) \\
$i$ & 111,357,023 & (2,007,214,004) & 23,860,632 & (430,025,357) & 23,853,509 & (429,848,423) & 159,071,164 & (2,867,087,784) \\
$z$ & 69,509,568 & (1,096,348,579) & 14,895,044 & (234,837,927) & 14,892,951 & (234,946,327) & 99,297,563 & (1,566,132,833) \\
$y$ & 43,478,036 & (639,187,709) & 9,318,969 & (137,009,523) & 9,313,135 & (136,956,295) & 62,110,140 & (913,153,527) \\
\hline
Total & 305,060,826 & (5,546,550,265) & 65,376,311 & (1,188,448,587) & 65,352,362 & (1,188,230,085) & 435,789,499 & (7,923,228,937) \\
\enddata
\tablecomments{As Table~\ref{tab:dp1-counts}, but for DP2, with split boundaries 0.7 and 0.85. The number of sources is given in parentheses.}
\end{deluxetable}

\begin{deluxetable}{lr@{\;}rr@{\;}rr@{\;}rr@{\;}r}
\tablecaption{Trainable single-band light curves and sources in DP2 difference-image forced source photometry.\label{tab:dp2-counts-diff}}
\tablehead{\colhead{Band} & \multicolumn{2}{c}{Train} & \multicolumn{2}{c}{Validation} & \multicolumn{2}{c}{Test} & \multicolumn{2}{c}{Total}}
\startdata
$u$ & 3,544,736 & (182,520,473) & 759,367 & (39,107,209) & 758,512 & (39,092,308) & 5,062,615 & (260,719,990) \\
$g$ & 15,441,642 & (545,957,631) & 3,309,530 & (117,149,202) & 3,306,973 & (116,991,155) & 22,058,145 & (780,097,988) \\
$r$ & 65,394,222 & (1,290,496,786) & 14,019,786 & (276,443,726) & 14,008,642 & (276,393,781) & 93,422,650 & (1,843,334,293) \\
$i$ & 117,244,491 & (2,245,767,220) & 25,127,264 & (481,297,189) & 25,115,995 & (481,089,382) & 167,487,750 & (3,208,153,791) \\
$z$ & 73,205,168 & (1,237,281,801) & 15,686,060 & (265,032,166) & 15,681,080 & (265,068,937) & 104,572,308 & (1,767,382,904) \\
$y$ & 45,754,979 & (684,845,560) & 9,805,105 & (146,785,781) & 9,801,349 & (146,758,838) & 65,361,433 & (978,390,179) \\
\hline
Total & 320,585,238 & (6,186,869,471) & 68,707,112 & (1,325,815,273) & 68,672,551 & (1,325,394,401) & 457,964,901 & (8,838,079,145) \\
\enddata
\tablecomments{As Table~\ref{tab:dp2-counts}, but for the forced PSF photometry on the difference images. The number of sources is given in parentheses.}
\end{deluxetable}

\subsection{Corrected uncertainties in the DP2 catalogs}\label{sec:data-corrected}

We added the corrected uncertainties obtained in this work to the DP2 HATS object and DIA object catalogs as new nested columns of the forced sources, \texttt{objectForcedSource} and \texttt{diaObjectForcedSource}, respectively\footnote{\url{https://lsdb.io/dp2}}.
The column \texttt{psfFluxErr\_corrected} contains the corrected uncertainties of the direct-image PSF fluxes (Section~\ref{sec:results-dp2}), and \texttt{psfDiffFluxErr\_corrected} contains those of the difference-image PSF fluxes (Section~\ref{sec:results-dp2-diff}).
The column \texttt{psfMagErr\_corrected} gives the corresponding magnitude errors of the direct-image photometry, derived in the same way as the other magnitude errors of the catalogs.
The flags \texttt{psfFluxErr\_corrected\_flag} and \texttt{psfDiffFluxErr\_corrected\_flag} are set when the correction factor $u(\theta_{ki})$ (Section~\ref{sec:models}) is outside the interval $[0.1, 50]$.
For the DIA object catalog, we apply the models trained on the object catalog as is.
The forced photometry at the DIA object positions may have different properties from that at the object positions.
However, DIA objects are variable by definition, so they cannot be used for training with our method, which relies on the light curves of non-variable objects, and transferring the models trained on the objects is the best we can do within this framework.
The corrected uncertainties may be used, for example, for light curve fitting and variability detection.

\section{Methods}\label{sec:methods}

In this section, we describe how we validate and correct the reported uncertainties.
We assume that the measurements of a non-variable object are independent and normally distributed around its constant flux, and that the reported uncertainties may be miscalibrated by a factor that depends on the properties of each measurement.
We first introduce the light curve whitening (Section~\ref{sec:whitening-motivation}), then the uncertainty correction models and their input features (Sections~\ref{sec:models} and~\ref{sec:features}), the loss functions used to fit them (Section~\ref{sec:losses}), the rejection of variable objects (Section~\ref{sec:rejection-of-variable-objects}), and the training procedure (Section~\ref{sec:training}).

\subsection{Light curve whitening}\label{sec:whitening-motivation}

Let $k$ be the light curve index, and $i$ be the photometric measurement index for a given light curve.
The $k$-th light curve has $N_k$ observations.
Then $f_{ki}$ is the measured flux, and $\sigma_{ki}$ is the corresponding uncertainty value given by the survey, both in the same units.

The weighted mean flux would be:
\begin{equation}\label{eq:average-sums}
\average{f_k} = \frac{\sum_{i=1}^{N_k}{f_{ki} / \sigma_{ki}^2}
}{\sum_{i=1}^{N_k}{1/\sigma_{ki}^2}}\,.
\end{equation}

We introduce the relative residual flux:
\begin{equation}\label{eq:rel-res-sums}
r_{ki} \equiv \frac{f_{ki} - \average{f_k}}{\sigma_{ki}}\,.
\end{equation}

The null hypothesis is that the survey's flux measurements for non-variable objects are sampled from $f_{ki}~\sim~N(\mu_k, \sigma_{ki}^2)$, where $\mu_k$ is the ``true'' flux of the object, which is unknown, and $\sigma_{ki}$ are known measurement uncertainties; all $f_{ki}$ are assumed to be independent.
Under this hypothesis, the relative residual flux vector for light curve $k$, $\{r_i\}|_k$, is normally, but not standard normally, distributed (though in the large-number limit it approximately is; see Appendices~\ref{sec:light-curve-whitening} and~\ref{sec:alternatives-variance}).

Next, we introduce $z_{kj}~\sim~N(0, 1)$\footnote{
Within the scope of this research, we assume that there is no correlation between observations of different light curves, so all components of $z_{kj}$ are independent.},
which are the components of vectors representing the whitened light curve signal; see Appendix~\ref{sec:light-curve-whitening} for the whitening procedure details.
A whitened signal vector $\{z_j\}|_k$ represents a single light curve, and, under the same hypothesis, it is distributed as the multidimensional standard normal distribution.
Notably, $\{z_j\}|_k$ has one fewer component than the corresponding $\{f_i\}|_k$ vector, $N_k-1$, because one degree of freedom vanishes due to the unknown $\mu_k$ value; we use the $j$ index for $z$ instead of $i$ to highlight that difference.

If we assume that the reported flux variations are sampled with a standard deviation different from the reported one, for example $u \cdot \sigma_{ki}$, where $u$ is the constant uncertainty underestimation factor, then the whitened signal $z_{kj}$ derived with the use of the reported $\sigma_{ki}$ uncertainties would follow $N(0, u^2)$ (see Eq.~\ref{eq:whitening-operator}).
That means that measuring the distribution of $\{z_{kj}\}$ gives us a proxy for how much the catalog error values are underestimated, for instance, in this specific case of constant $u$, the variance of the distribution would be enough to quantify the uncertainty correction.
In a more general case when the uncertainty underestimation may depend on each individual source metadata, we should find a way to correct each individual uncertainty in a way that makes $\{z_{kj}\}$ follow the standard normal distribution.
Practically, we would define $u$ as a function of source parameters, such as reported flux, flux error, and, optionally, observational metadata, e.g. passband, seeing, sky background or distance of the source from the center of the field of view.

\subsection{Uncertainty correction models}\label{sec:models}

We parameterize the correction as a function of the source features, $u(\theta_{ki})$, so that the corrected uncertainty of the $i$-th measurement of the $k$-th light curve is $u(\theta_{ki})\,\sigma_{ki}$; the features $\theta$ are described in Section~\ref{sec:features}.
We consider two models of different complexity.
The first one is a multilayer perceptron (MLP), a flexible model which may capture the dependence of the correction on all the features we provide.
The second one is a constant magnitude error added in quadrature to the reported uncertainty, with a single parameter per passband, which has a simple analytical form.
Both models are trained with the same pipeline, so comparing them indicates which part of the miscalibration may be described by a simple error floor.

\paragraph{Multilayer perceptron}
The MLP takes the features $\theta$, scaled as described in Section~\ref{sec:features}, as the input, and has three fully connected hidden layers of 300, 300, and 400 neurons with the Gaussian error linear unit (GELU) activation function \citep{2016arXiv160608415H}, followed by a linear output layer with a single output $y(\theta)$.
The uncertainty underestimation factor is then
\begin{equation}\label{eq:mlp-u}
    u(\theta) = \exp\left(-y(\theta)\right)\,,
\end{equation}
which keeps $u$ positive and allows both underestimated ($u > 1$) and overestimated ($u < 1$) uncertainties.
We train a single MLP for all passbands, with the passband provided as a one-hot encoded feature\footnote{For example, the $r$ band is $(0, 0, 1, 0, 0, 0)$, with the components ordered as $ugrizy$.}, and a separate MLP for each data preview and photometry type.
We do not use dropout or any regularization of the network weights, besides the penalty term of the loss function (Eq.~\ref{eq:penalty}).
The MLP has no simple analytical form, so we study the contribution of the individual features to its output with the feature importance analysis (Section~\ref{sec:discussion-features}).

\paragraph{Per-band constant magnitude error}
A simple way to account for the errors not included in the reported uncertainties, for example, photometric calibration errors, is to add a constant ``error floor'' to the magnitude uncertainty in quadrature.
The reported magnitude uncertainty, obtained from the flux uncertainty by error propagation, is $\sigma_{m,ki} = 2.5 / \ln{10} \cdot \,\sigma_{ki} / F_{ki}$, where $F_{ki}$ is the flux of the source.
For the direct-image photometry, $F_{ki} = f_{ki}$; for the difference-image photometry, we use the forced PSF flux measured on the calibrated (direct) image of the same visit as $F_{ki}$, because the noise associated with the source flux is still present on the difference image, while $f_{ki}$ is consistent with zero for a non-variable object.
The corrected magnitude uncertainty is $\sqrt{\sigma_{m,ki}^2 + \Delta m_\mathrm{band}^2}$, where $\Delta m_\mathrm{band}$ is the trainable additive magnitude error for the passband of the light curve, which gives
\begin{equation}\label{eq:const-magerr-u}
    u(F_{ki}, \sigma_{ki}, \mathrm{band}) = \sqrt{1 + \left(\frac{\Delta m_\mathrm{band}}{\sigma_{m,ki}}\right)^2} = \sqrt{1 + \left(\frac{\ln 10}{2.5} \frac{\Delta m_\mathrm{band}\,F_{ki}}{\sigma_{ki}}\right)^2}\,.
\end{equation}
In terms of fluxes, the model adds a term proportional to the source flux to the reported flux uncertainty in quadrature.

\subsection{MLP input features}\label{sec:features}

The input features $\theta$ of the MLP model (Section~\ref{sec:models}) combine per-source measurements, per-object properties, and per-visit observational parameters.
All features are scaled to a roughly unit range before being passed to the MLP.
Table~\ref{tab:features} lists the features and their scaling.

\begin{deluxetable}{llll}
\tablecaption{Input features of the MLP models.\label{tab:features}}
\tablehead{\colhead{Feature} & \colhead{Catalog column} & \colhead{Level} & \colhead{Scaling}}
\startdata
Source PSF flux $f$, nJy & \texttt{psfFlux} / \texttt{psfDiffFlux} & source & $\arsinh$ \\
Source PSF flux error $\sigma$, nJy & \texttt{psfFluxErr} / \texttt{psfDiffFluxErr} & source & $\log$ \\
Direct-image PSF flux\tablenotemark{$\dagger$}, nJy & \texttt{psfFlux} & source & $\arsinh$ \\
Extendedness & \texttt{<band>\_extendedness} & object & none \\
Exposure time $t_\mathrm{exp}$, s & \texttt{expTime} & visit & $\log$ \\
Sky background $B_\mathrm{sky}$, ADU & \texttt{skyBg} & visit & $\log$ \\
Seeing, arcsec & \texttt{seeing} & visit & $\log$ \\
Detector position $(\rho, \cos\phi, \sin\phi)$ & \texttt{detector} & visit & none\tablenotemark{$\ddagger$} \\
Passband & \texttt{band} & source & one-hot \\
\enddata
\tablenotetext{\noexpand\ensuremath{\noexpand\dagger}}{Difference-image photometry only: the forced PSF flux measured on the calibrated (direct) image of the same visit.}
\tablenotetext{\noexpand\ensuremath{\noexpand\ddagger}}{Used as is; the detector position coordinates are described in the text.}
\end{deluxetable}

The source PSF flux $f$, given in nJy, is scaled as
\begin{equation}
\frac{\arsinh(f / f_\mathrm{low})}{\arsinh(f_\mathrm{high} / f_\mathrm{low})}\,,
\end{equation}
where $f_\mathrm{low}$ and $f_\mathrm{high}$ correspond to AB magnitudes 23 and 14, which roughly bound the magnitude range of the survey.
This scaling behaves like the asinh magnitudes \citep{1999AJ....118.1406L}: it is logarithmic for bright fluxes, while remaining linear and well-defined around zero and for negative fluxes, which are common for faint sources and difference-image photometry.
The reported flux error $\sigma$, also in nJy, is scaled logarithmically with min--max normalization, $\tilde{\sigma} = (\lg\sigma + 1) / 5$, corresponding to the range $\lg\sigma \in [-1, 4]$.

We also use the object extendedness in the same passband as the light curve; note that the catalogs provide one value per band, along with a reference-band one.
Extendedness is a binary star/galaxy classifier based on the comparison of fluxes measured with two models: it is set to 1 if the PSF model flux is less than $98.5\%$ of the free (not forced) composite-model (CModel) flux in the given band, i.e. if the PSF magnitude exceeds the CModel magnitude by more than 16\,mmag, and to 0 otherwise \citep{2026AJ....171..360V}.

The per-visit parameters are the exposure time, scaled as $\lg(t_\mathrm{exp} / 30\,\mathrm{s})$, the sky background in analog-to-digital units (ADU), scaled as $\lg(B_\mathrm{sky} / 10^3\,\mathrm{ADU})$, and the seeing, scaled as $\lg(\mathrm{seeing} / 1\,\mathrm{arcsec})$.
The position of the detector center in the focal plane is encoded in polar coordinates as $(\rho, \cos\phi, \sin\phi)$, with the origin at the center of the focal plane; this encoding reflects the approximate radial symmetry of optical distortions.
For DP1, LSSTComCam's nine detectors are placed on a unit-spaced $3\times3$ grid, so $\rho \in \{0, 1, \sqrt{2}\}$; for DP2, we use the actual focal-plane positions of the 189 LSSTCam science detector centers from the camera geometry model.
Finally, the band is one-hot encoded.

The trained MLP models are distributed with the \texttt{uncle-val} package (Section~\ref{sec:implementation}) as PyTorch and Open Neural Network Exchange (ONNX) files with the feature scaling embedded, so they take the unscaled feature values of Table~\ref{tab:features} as input.

\subsection{Loss functions}\label{sec:losses}

Throughout this section, $u(\theta)$ is the uncertainty underestimation factor predicted by a model (Section~\ref{sec:models}) from the source features $\theta$.

The models are trained on batches of $B$ light curves, each represented by $\Nobs = 10$ randomly selected sources in random order; we index light curves within a batch by $b = 1 \dots B$.
For every light curve $b$, the fluxes and the corrected uncertainties $u(\theta_{bi}) \cdot \sigma_{bi}$ are whitened into $z_{bj}$, $j = 1 \dots \Nobs - 1$; as discussed in Section~\ref{sec:whitening-motivation} and Appendix~\ref{sec:light-curve-whitening}, one degree of freedom per light curve is lost to the unknown mean flux, so a batch yields $B(\Nobs - 1)$ whitened values in total.
Under the null hypothesis of correct uncertainties, all $z_{bj}$ are independent samples from $N(0, 1)$, so each loss function below is a differentiable measure of the departure of the whitened sample from standard normality.

The first loss function is based on the $\chi^2$ statistic of the relative residuals (Eq.~\ref{eq:rel-res-sums}), $\chi^2_b \equiv \sum_i r_{bi}^2$.
Under the null hypothesis, the batch sum follows the $\chi^2$ distribution with $\nu = B(\Nobs - 1)$ degrees of freedom.
The loss is the negative log-likelihood:
\begin{equation}\label{eq:chi2-loss}
    \mathcal{L}_{\chi^2} = -\ln p_{\chi^2_\nu}\left( \sum_{b=1}^{B} \chi^2_b \right)\,,
\end{equation}
where $p_{\chi^2_\nu}$ is the probability density function of the $\chi^2_\nu$ distribution.

The other two loss functions are built on the whitened values $z_{bj}$ of all light curves of the batch.
To bound the influence of extreme outliers, such as unfiltered variables and image artifacts, the whitened values are softened before computing the loss:
\begin{equation}\label{eq:softening}
    \tilde{z} = a \tanh\left(z / a\right)\,,
\end{equation}
where $a$ is the softening scale, for which we use $a = 20$; the transformation is close to identity for $|z| \ll a$ and smoothly saturates at $\pm a$.
For the $\chi^2$ loss, the same transformation is applied to the relative residuals $r_{bi}$ instead.

The second loss function is the Kullback--Leibler (KL) divergence between a normal distribution fitted to the whitened sample and the standard normal distribution:
\begin{equation}\label{eq:kl-loss}
    \mathcal{L}_\mathrm{KL} = \mathrm{KL}\left(N(\hat\mu_z, \hat\sigma_z^2)\,\|\,N(0,1)\right) = \frac12 \left( \hat\mu_z^2 + \hat\sigma_z^2 - \ln \hat\sigma_z^2 - 1 \right)\,,
\end{equation}
where $\hat\mu_z$ and $\hat\sigma_z^2$ are the sample mean and the unbiased sample variance of all $B(\Nobs-1)$ whitened values of the batch.
This loss matches the first two moments of the whitened sample to those of the standard normal distribution.

The third loss function, which we use as the main training loss, is based on the Epps--Pulley statistic \citep{epps1983test}: a weighted $L^2$ distance between the empirical characteristic function of the whitened sample and that of the standard normal distribution.
Let $\omega$ be the frequency variable of the characteristic functions, $\varphi(\omega) = \exp\left(-\omega^2/2\right)$ be the characteristic function of the standard normal distribution, and $\hat\varphi(\omega) = \average{\exp\left(\mathrm{i} \omega z\right)}$ be the empirical characteristic function, with the average taken over all whitened values of the batch.
Then the loss is
\begin{equation}\label{eq:ep-loss}
    \mathcal{L}_\mathrm{EP} = B(\Nobs - 1) \int_{-\infty}^{+\infty} \left| \hat\varphi(\omega) - \varphi(\omega) \right|^2 w(\omega) \,\mathrm{d}\omega\,,
\end{equation}
where we take the weight function equal to the standard normal characteristic function, $w(\omega) = \varphi(\omega)$ \citep{2025arXiv251108544B}, and evaluate the integral with the trapezoidal rule on a uniform grid of 17 nodes over $\omega \in [-5, 5]$.
Unlike the moment-based losses, this loss is sensitive to the whole shape of the distribution, while remaining smooth and cheap to evaluate.

Finally, a penalty term may be added to any of the losses:
\begin{equation}\label{eq:penalty}
    \mathcal{L} = \mathcal{L}_\mathrm{stat} + \lambda \, \gamma \, \average{u(\theta)}\,,
\end{equation}
where $\mathcal{L}_\mathrm{stat}$ is any of the three losses above, $\lambda$ is the penalty strength, the average is taken over all sources of the batch, and $\gamma$ is a loss-dependent scale factor: $\gamma = 1$ for the KL and Epps--Pulley losses, and $\gamma = |\ln p_{\chi^2_\nu}(\nu)|$ for the $\chi^2$ loss.
The penalty discourages large values of $u$, so that the model inflates the reported uncertainties only as much as the data require.
In our fiducial setup, we train with the Epps--Pulley loss with $\lambda = 2$ and track the other two losses (without the penalty) as validation metrics.
We compare the models trained with the different loss functions in Section~\ref{sec:discussion-losses}.

\subsection{Rejection of variable objects}\label{sec:rejection-of-variable-objects}

Since the analysis relies on the assumption of constant flux, we reject variable objects using four complementary variability criteria, which are applied after the source flag filtering (Section~\ref{sec:data}).
Each single-band light curve is tested separately and removed if any of the criteria is satisfied.
All criteria operate on a single light curve, so throughout this section we consider a fixed light curve $k$ and drop this index: $f_i$ and $\sigma_i$, $i = 1 \dots N$, denote the fluxes and the reported uncertainties, and $N$ is the number of observations.
We denote the median flux as $f_\mathrm{med}$.

\paragraph{Large magnitude amplitude} The first criterion targets large-amplitude variables, such as eclipsing binaries and long-period pulsating stars.
Among the observations with positive flux and magnitude error below 50\,mmag (i.e., $\frac{2.5}{\ln 10}\,\sigma_i / f_i < 0.05$), the light curve is marked as variable if its amplitude exceeds 1\,mag.

\paragraph{Large flux deviation from the median} This criterion catches strong flux outliers, such as image artifacts and transient contamination: the light curve is marked as variable if any observation deviates from the median flux by more than $100\,\sigma_i$:
\begin{equation}\label{eq:flux-outlier}
    \left| f_i - f_\mathrm{med} \right| > 100\,\sigma_i\,.
\end{equation}

\paragraph{Large deviation of the residual} This is a self-normalized version of the previous one.
We introduce the median deviations
\begin{equation}\label{eq:median-deviation}
    g_i \equiv \frac{f_i - f_\mathrm{med}}{\sigma_i}\,,
\end{equation}
and mark the light curve as variable if any $|g_i|$ exceeds five sample standard deviations of $\{g_i\}$.
Normalizing the threshold by the scatter of the deviations themselves, rather than by the reported uncertainties, keeps this criterion meaningful even when the uncertainties are miscalibrated, which is the working hypothesis of this paper.

\paragraph{Lag-one autocorrelation} It catches slow, low-amplitude variables, which produce correlated consecutive observations rather than outliers.
With time stamps ignored, we compute the lag-one serial correlation coefficient of the fluxes about the median \citep{anderson1942serial}:
\begin{equation}\label{eq:autocorrelation}
    r_1 = \frac{\sum_{i=1}^{N-1} \left(f_i - f_\mathrm{med}\right)\left(f_{i+1} - f_\mathrm{med}\right)}{\sum_{i=1}^{N} \left(f_i - f_\mathrm{med}\right)^2}\,,
\end{equation}
and mark the light curve as variable if $|r_1| > 5 / N$.

\begin{figure*}
    \centering
    \includegraphics[width=0.49\textwidth]{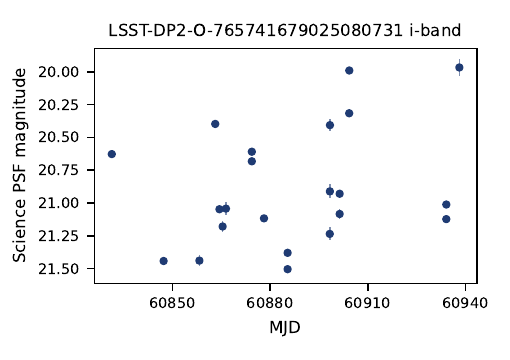}\hfill
    \includegraphics[width=0.49\textwidth]{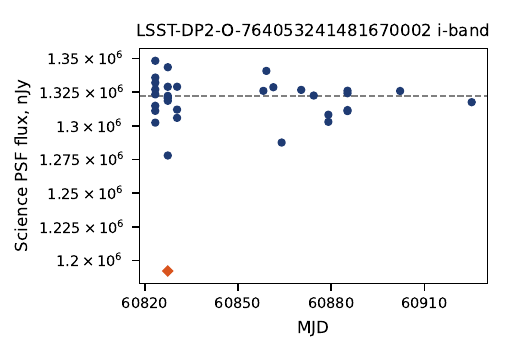}\\
    \includegraphics[width=0.49\textwidth]{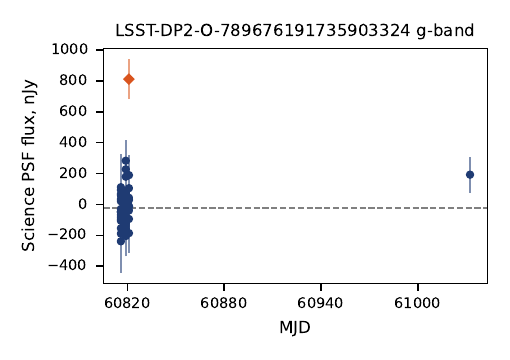}\hfill
    \includegraphics[width=0.49\textwidth]{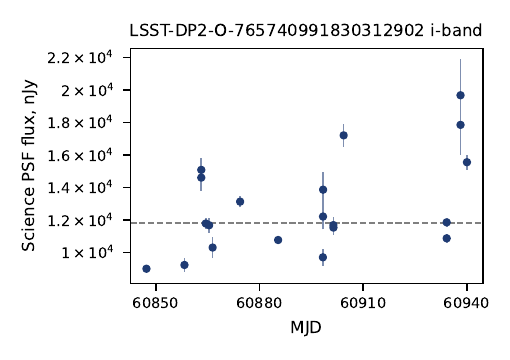}
    \caption{Examples of DP2 direct-image light curves, each rejected by exactly one of the variability criteria: large magnitude amplitude (top left), large flux deviation from the median (top right), large deviation of the residual (bottom left), and lag-one autocorrelation (bottom right). Dashed lines show the median flux $f_\mathrm{med}$, and diamonds mark the sources that satisfy the outlier criteria in the two cases where specific sources can be identified as the reason for rejection.\label{fig:dp2-variable-examples}}
\end{figure*}

The criteria listed above would not perfectly clean the dataset of the variable light curves, but would be sufficient to remove objects with large variability and photometric outliers.
We assume that the combination of this rejection and softening of the loss function (see Section~\ref{sec:losses}) would be enough to make the contribution of the variability signals negligible for the ground-based photometric data.
Figure~\ref{fig:dp2-variable-examples} shows examples of DP2 light curves, each rejected by exactly one of these criteria.

\subsection{Training and implementation}\label{sec:training}\label{sec:implementation}

All the analysis is implemented in the open-source \texttt{uncle-val} Python package \citep{malanchev2026uncleval}\footnote{\url{https://github.com/lincc-frameworks/uncle-val}}, which also distributes the trained MLP models (Section~\ref{sec:features}).
We process the catalogs with \texttt{LSDB} \citep{2025arXiv250102103C} and stream the light curves from the catalog partitions into PyTorch, so the training dataset is never materialized: the quality cuts and the rejection of variable objects (Section~\ref{sec:rejection-of-variable-objects}) are applied on the fly, in parallel for each partition.

For the same reason, we split the objects into the training, validation, and test sets on the fly, by the hash of the object ID, so the split is reproducible and all light curves of an object belong to the same set.
The hash is the FarmHash fingerprint normalized to $[0, 1)$, and the split boundaries are given in Section~\ref{sec:data}.
We use only light curves with at least $\Nobs = 10$ sources.
Each time a training light curve is drawn, we take $\Nobs$ of its sources at random without replacement; validation light curves are subsampled the same way but only once, and test light curves are used in full.

We use the Adam optimizer \citep{kingma2015adam} and evaluate the validation loss after every portion of training light curves twice the size of the validation set.
The learning rate is reduced by a factor of $10^{1/2}$ after 16 evaluations without improvement, and the model with the lowest validation loss is selected as the final one.
The MLPs are trained with an initial learning rate of $10^{-4}$ on batches of 512 light curves, for $3 \times 10^8$ light curves in total, and validated on 131,072 light curves.

We train the per-band constant magnitude error model on the DP2 difference-image photometry only, separately for each of the three fields: Rubin-SV-225 and two Deep Drilling Fields (DDF), DDF-COSMOS and DDF-EDFS, the latter covering the Euclid Deep Field South (EDFS).
For this model, we also select only point sources (extendedness${}=0$) brighter than 21\,mag in the band of the light curve.
The model has only six parameters, so we use a larger initial learning rate of $10^{-3}$ and batches of 1024 light curves, for $5 \times 10^6$ light curves in total.

\section{Results}\label{sec:results}

We first present in detail the MLP model for the DP2 direct-image forced source photometry, which is our primary result (Section~\ref{sec:results-dp2}).
Then we compare it, more briefly, with the DP2 difference-image forced source photometry (Section~\ref{sec:results-dp2-diff}), with the per-band constant magnitude error model (Section~\ref{sec:results-const-magerr}), and with DP1 (Section~\ref{sec:results-dp1}).
All results are obtained on the test set, using all sources of each light curve (Section~\ref{sec:training}).

\subsection{DP2 direct-image photometry}\label{sec:results-dp2}

\begin{figure*}
    \centering
    \includegraphics[width=\textwidth]{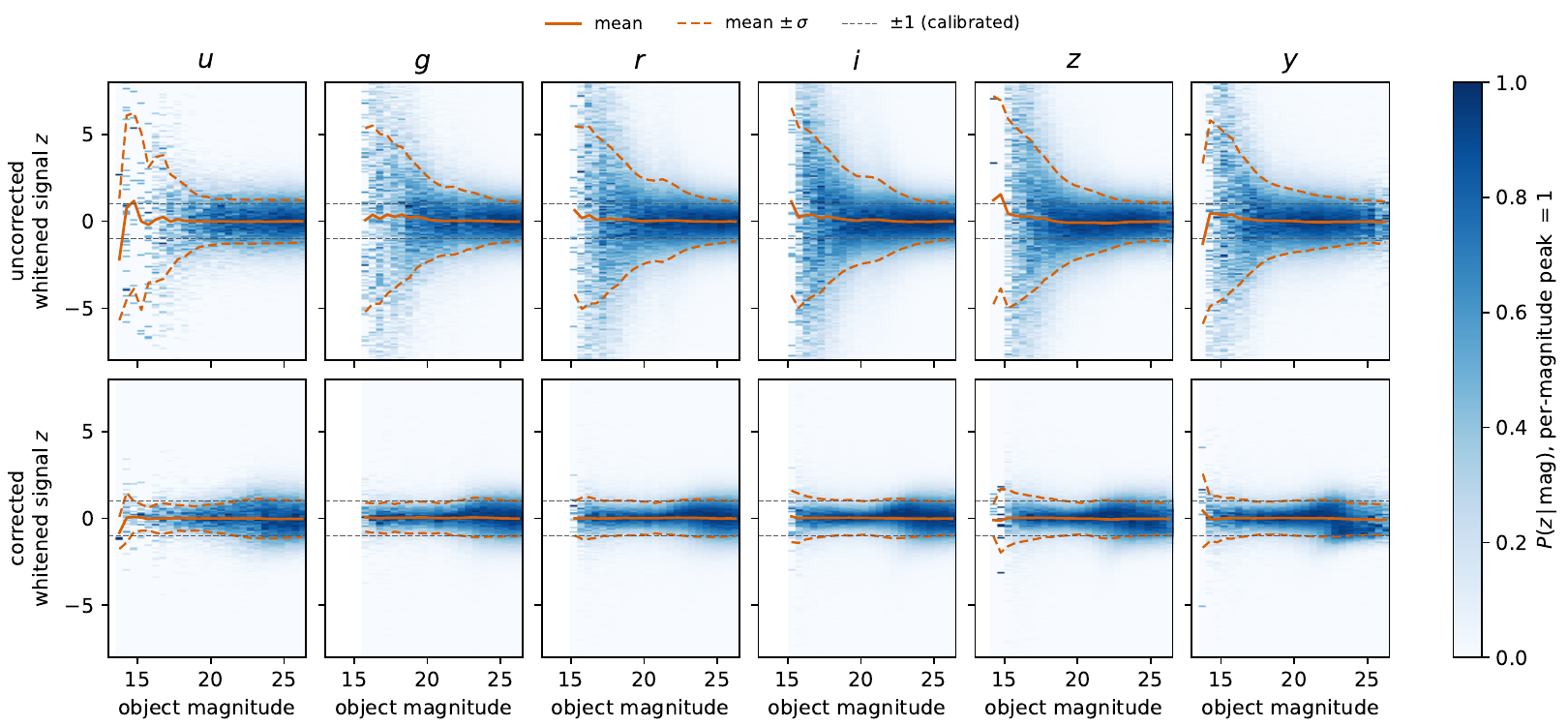}
    \caption{Per-band density of the whitened signal $z$ as a function of object PSF magnitude for the DP2 direct-image forced source photometry (test set), with the reported (top row) and the MLP-corrected (bottom row) uncertainties. Every magnitude column is normalized to its own peak. The solid curve is the mean, the dashed curves are the mean $\pm$ the standard deviation, and the thin dashed lines mark $\pm 1$.\label{fig:dp2-direct-z}}
\end{figure*}

Figure~\ref{fig:dp2-direct-z} shows the distribution of the whitened signal $z$ as a function of the object PSF magnitude in the same band, for each band separately.
Each column of a panel is the histogram of $z$ for the light curves in a narrow magnitude bin, normalized to its peak, so the color shows the shape of the conditional distribution $P(z \,|\, \mathrm{mag})$ independently of the number of objects in the bin.
The solid curve is the mean of $z$, and the dashed curves are the mean plus and minus the standard deviation of $z$; for correct uncertainties they follow 0 and $\pm 1$, the latter shown by the thin dashed lines.
Since a multiplicative change of the uncertainties $\sigma \to u \cdot \sigma$ changes the whitened signal as $z \to z / u$ (Appendix~\ref{sec:light-curve-whitening}), the standard deviation of $z$ in a magnitude bin estimates the typical uncertainty underestimation factor of the sources in this bin.

The top row of Figure~\ref{fig:dp2-direct-z} shows $z$ obtained with the reported uncertainties.
In the $u$-band, the standard deviation of $z$ is larger than unity over the whole magnitude range, while in the other bands it exceeds unity only for objects brighter than about 22\,mag, so the uncertainties are underestimated mostly for bright sources.
The underestimation strongly depends on the magnitude: the standard deviation of $z$ grows towards the bright end, reaching about 5 for objects brighter than 17--18\,mag.
This behavior is consistent with an unaccounted error correlated with the source flux, such as a calibration or a PSF modeling error.

The bottom row of Figure~\ref{fig:dp2-direct-z} shows $z$ obtained with the corrected uncertainties $u(\theta_{ki})\,\sigma_{ki}$.
The correction brings the standard deviation of $z$ close to unity over the whole magnitude range in all bands, with small deviations only for the brightest objects.
For objects brighter than about 18\,mag, the correction corresponds to an additional magnitude error of about 10\,mmag added in quadrature to the reported one.

\begin{figure*}
    \centering
    \includegraphics[width=0.49\textwidth]{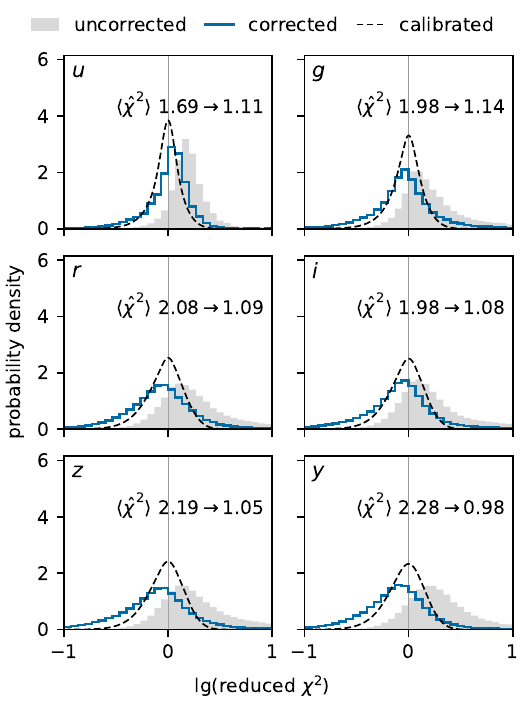}\hfill
    \includegraphics[width=0.49\textwidth]{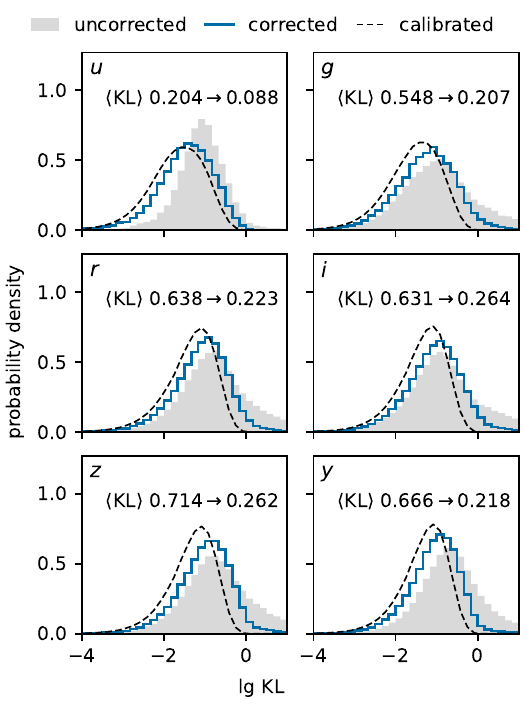}
    \caption{Per-band distributions of the reduced $\chi^2$ (left) and of the KL divergence (right) of the light curves for the DP2 direct-image forced source photometry (test set), with the reported (filled) and the MLP-corrected (line) uncertainties. The dashed curves are the distributions for correct uncertainties.\label{fig:dp2-direct-metrics}}
\end{figure*}

Figure~\ref{fig:dp2-direct-metrics} checks the correction with two other metrics, computed per light curve using all its $N_k$ sources.
The first metric is the reduced $\chi^2$,
\begin{equation}\label{eq:reduced-chi2}
    \hat\chi^2_k = \frac{1}{N_k - 1} \sum_{j=1}^{N_k - 1} z_{kj}^2 = \frac{1}{N_k - 1} \sum_{i=1}^{N_k} r_{ki}^2\,.
\end{equation}
For correct uncertainties, $(N_k - 1)\,\hat\chi^2_k \sim \chi^2_{N_k - 1}$.
Since $N_k$ differs between light curves, the reference distribution is the mixture over the observed distribution of $N_k$,
\begin{equation}\label{eq:reduced-chi2-cal}
    p_\mathrm{cal}\!\left(\hat\chi^2\right)
    = \sum_{N} w_N \, p_{\chi^2_{N-1}/(N-1)}\!\left(\hat\chi^2\right)\,,
\end{equation}
where $p_{\chi^2_{N-1}/(N-1)}$ is the density of a $\chi^2_{N-1}$ variable divided by $N-1$, and $w_N$ is the fraction of light curves with $N_k = N$.
Underestimated uncertainties shift the distribution of $\hat\chi^2_k$ towards values larger than unity.
With the reported uncertainties, the mean $\hat\chi^2_k$ is 1.7--2.3, depending on the band (left panel of Figure~\ref{fig:dp2-direct-metrics}).
The correction brings it to 0.98--1.14 and centers the distribution on unity, but the distribution remains broader than the reference one, with excess light curves on both sides.

The second metric is the KL divergence of Eq.~\ref{eq:kl-loss}, computed with the mean and the variance of $z_{kj}$ of a single light curve.
Its distribution for correct uncertainties has no simple closed form, so we compute it numerically: for each light curve length $N$, we draw $N-1$ independent standard normal values many times, compute the KL divergence of each draw, and average the resulting distributions with the weights $w_N$ of Eq.~\ref{eq:reduced-chi2-cal}.
The correction reduces the mean KL divergence from 0.20--0.71 to 0.09--0.26, depending on the band (right panel of Figure~\ref{fig:dp2-direct-metrics}), but, unlike the reduced $\chi^2$, the corrected distribution remains shifted towards larger values than the reference one.

\subsection{DP2 difference-image photometry}\label{sec:results-dp2-diff}

\begin{figure*}
    \centering
    \includegraphics[width=0.49\textwidth]{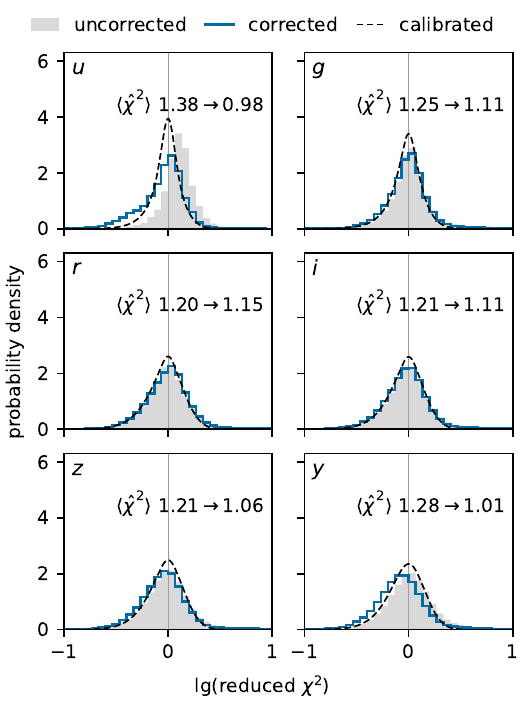}\hfill
    \includegraphics[width=0.49\textwidth]{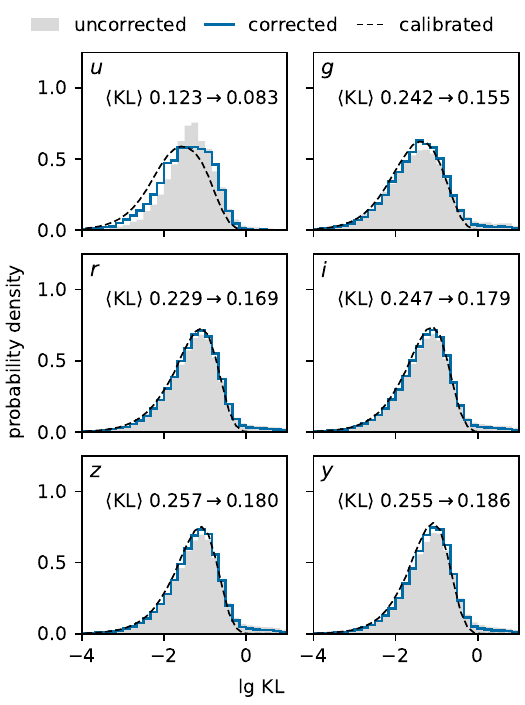}
    \caption{Same as Figure~\ref{fig:dp2-direct-metrics}, for the DP2 difference-image forced source photometry.\label{fig:dp2-diff}}
\end{figure*}

Per light curve, the reported uncertainties of the DP2 difference-image forced source photometry are much closer to correct than those of the direct-image photometry (Figure~\ref{fig:dp2-diff}).
With the reported uncertainties, the mean reduced $\chi^2$ is 1.20--1.38 and the mean KL divergence is 0.12--0.26, depending on the band; the correction changes them to 0.98--1.15 and 0.08--0.19, respectively.
Except for the $u$-band, the distributions of both statistics are close to the reference ones before and after the correction.

We understand that the corrected uncertainties of the difference-image photometry, which we provide in the DP2 catalogs (Section~\ref{sec:data-corrected}), are most interesting for variable objects, which our training set excludes by construction.
For a non-variable object, the difference flux is consistent with zero, while for a variable object it is not, so the uncertainties of the latter may include additional terms that scale with the difference flux, such as those caused by errors of the PSF model and of the image subtraction.
Therefore, for variable objects, we consider the corrected difference-image uncertainties to be lower limits on the true ones.

\subsection{Per-band constant magnitude error}\label{sec:results-const-magerr}

We compare the MLP with the per-band constant magnitude error model (Eq.~\ref{eq:const-magerr-u}), which adds a single magnitude error $\Delta m_\mathrm{band}$ in quadrature to the reported magnitude uncertainty in each band.
We fit this model to the DP2 difference-image forced source photometry with the same loss function as the MLP (Section~\ref{sec:losses}).
We use only point sources (extendedness${}=0$), to exclude extended objects, for which the PSF photometry is less reliable.
We also use only objects brighter than 21\,mag: for fainter objects, the reported magnitude uncertainty is much larger than $\Delta m_\mathrm{band}$, so $u \approx 1$ for them, and they carry little information about $\Delta m_\mathrm{band}$.

To see how the fitted error depends on the depth of the data, we fit the model separately for the three DP2 fields, which have different numbers of visits (Table~\ref{tab:const-magerr}).
The fitted error depends much more on the field than on the band: for example, in the $i$-band it decreases from about 11\,mmag in DDF-EDFS to about 7\,mmag in Rubin-SV-225.
In all bands, it grows from Rubin-SV-225 to DDF-COSMOS to DDF-EDFS, i.e.\ as the number of visits and the depth of the coadd decrease.
For the deepest field, Rubin-SV-225, the fitted error is 5.7--6.9\,mmag, almost the same in all bands.
This is similar to the photometric repeatability of bright stars in DP1, 5.1--7.1\,mmag in the six bands \citep{2026AJ....171..360V}.

\begin{deluxetable*}{lrrrrrrrrr}
\tablecaption{Fitted magnitude errors of the per-band constant magnitude error model for the DP2 difference-image forced source photometry.\label{tab:const-magerr}}
\tablehead{
    \colhead{} & \multicolumn{3}{c}{Rubin-SV-225} & \multicolumn{3}{c}{DDF-COSMOS} & \multicolumn{3}{c}{DDF-EDFS} \\
    \colhead{Band} & \colhead{$N_\mathrm{vis}$} & \colhead{$m_5$} & \colhead{$\Delta m$} & \colhead{$N_\mathrm{vis}$} & \colhead{$m_5$} & \colhead{$\Delta m$} & \colhead{$N_\mathrm{vis}$} & \colhead{$m_5$} & \colhead{$\Delta m$} \\
    \colhead{} & \colhead{} & \colhead{(mag)} & \colhead{(mmag)} & \colhead{} & \colhead{(mag)} & \colhead{(mmag)} & \colhead{} & \colhead{(mag)} & \colhead{(mmag)}
}
\startdata
$u$ & 312 & 26.26 & 6.3 & 105 & 25.77 & 9.3 & 7/5 & \nodata/23.69 & \nodata \\
$g$ & 529 & 27.26 & 5.9 & 87 & 26.73 & 7.5 & 31/37 & 25.63/25.75 & 14.6 \\
$r$ & 391 & 26.75 & 6.5 & 173 & 26.35 & 9.0 & 33/33 & 25.53/25.47 & 12.8 \\
$i$ & 381 & 26.36 & 6.9 & 162 & 25.94 & 8.7 & 59/59 & 25.65/25.71 & 11.4 \\
$z$ & 241 & 25.52 & 5.7 & 110 & 25.00 & 7.2 & 34/34 & 24.53/24.58 & 10.4 \\
$y$ & 104 & 24.06 & 6.5 & 77 & 23.69 & 8.0 & 9/10 & 23.08/23.27 & 10.7 \\
\enddata
\tablecomments{The model is trained on point sources (extendedness${}=0$) brighter than 21\,mag (Section~\ref{sec:training}). $N_\mathrm{vis}$ is the number of visits in the field, and $m_5$ is the $5\sigma$ PSF magnitude depth of the deep coadd at the center of the field \citep{2026lsst.data...42N}; for DDF-EDFS, they are given for its two parts, a/b. The $u$-band depth of DDF-EDFS~a is not available, and the $u$-band magnitude error for DDF-EDFS is not given because of insufficient data.}
\end{deluxetable*}

An imperfect chromatic calibration would make the error depend on the color of the object, so we check whether the fitted error depends on it.
For this test, we use all DP2 data together, with the same selection of point sources brighter than 21\,mag, and fit the model separately for blue objects, $0 \le g - r < 0.5$, and for red objects, $1.5 \le g - r < 2$, where $g - r$ is the difference of the object PSF magnitudes.
Without a color cut, the fitted error is 8.8--9.9\,mmag, depending on the band, and the fits for blue and red objects differ from it by at most 1.3\,mmag, which is small compared to the difference between the fields. Thus, we find no evidence of a color dependence of the fitted error.

\subsection{Comparison with DP1}\label{sec:results-dp1}

\begin{figure*}
    \centering
    \includegraphics[width=0.49\textwidth]{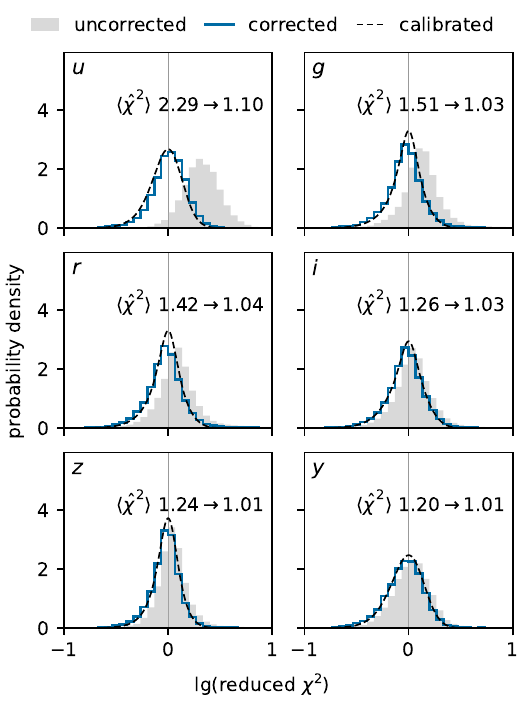}\hfill
    \includegraphics[width=0.49\textwidth]{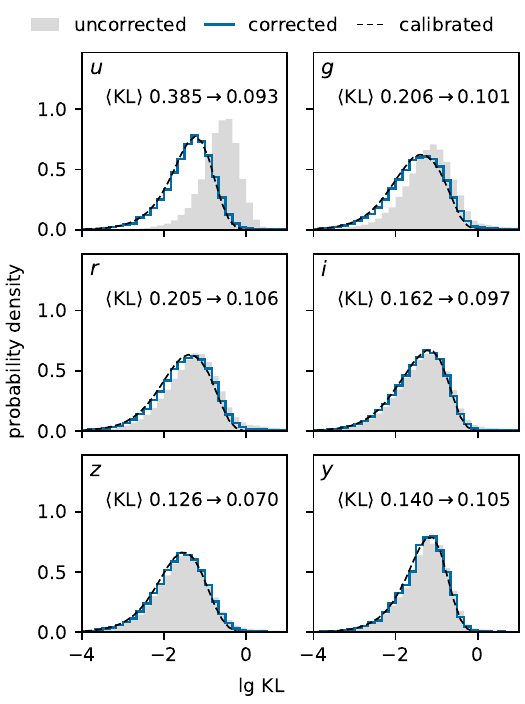}
    \caption{Same as Figure~\ref{fig:dp2-direct-metrics}, for the DP1 direct-image forced source photometry.\label{fig:dp1-metrics}}
\end{figure*}

Figure~\ref{fig:dp1-metrics} shows the reduced $\chi^2$ and the KL divergence for the MLP model trained on the DP1 direct-image forced source photometry.
With the reported uncertainties, the mean reduced $\chi^2$ is 1.20--2.29 and the mean KL divergence is 0.13--0.39, depending on the band, with the largest values in the $u$-band; the correction brings them to 1.01--1.10 and 0.07--0.11, respectively.
Unlike for DP2 (Figure~\ref{fig:dp2-direct-metrics}), the corrected distributions of both statistics match the reference ones in all bands.
Except for the $u$-band, the reported uncertainties of DP1 are also closer to correct than those of DP2, for which the mean reduced $\chi^2$ is 1.7--2.3.

We see two possible reasons why an MLP of the same architecture, trained in the same way, corrects the DP1 uncertainties better than the DP2 ones.
First, DP1 is much smaller: it has about 1.4 million trainable light curves (Section~\ref{sec:data}), about 300 times fewer than DP2, so a model of the same size may fit it better.
Second, the field of view of LSSTComCam is much smaller than that of LSSTCam, so the PSF varies less over it, and the photometry may be less affected by errors that depend on the position in the focal plane.

\section{Discussion}\label{sec:discussion}

\subsection{Feature importance}\label{sec:discussion-features}

We study the contribution of the input features to the MLP correction with SHapley Additive exPlanations \citep[SHAP;][]{lundberg2017shap}.
We explain the uncertainty underestimation factor $u(\theta)$ predicted by the MLP for individual sources.
For a source with features $\theta$, the prediction is decomposed into a sum of contributions of the individual features $\theta^{(m)}$:
\begin{equation}\label{eq:shap}
    u(\theta) = \average{u}_\mathrm{bg} + \sum_m \psi_m(\theta)\,,
\end{equation}
where $\average{u}_\mathrm{bg}$ is the mean prediction over a reference (background) sample of sources, and $\psi_m(\theta)$ is the SHAP value of the $m$-th feature.
For example, $\psi_m = 1$ means that, for this source, the value of the $m$-th feature increases the predicted $u$ by one compared to the mean prediction $\average{u}_\mathrm{bg}$, e.g., from 1.2 to 2.2, and $\psi_m = 0$ means that the feature does not change the prediction.
The SHAP values are based on the Shapley values of cooperative game theory \citep{shapley1953value}: the contribution of a feature is the change in the model output when the feature is set to its actual value, averaged over all possible orders in which the features are set, while the features not set yet take their values from the background sample.
We compute the SHAP values with the gradient-based approximation implemented in the \texttt{SHAP} package, known as expected gradients \citep{sundararajan2017axiomatic, 2021NatMI...3..620E}, using 500 background sources and explaining 2000 other sources of the test set.

As shown in Section~\ref{sec:results-dp2}, the magnitude of the source is the main factor of the uncertainty underestimation, which is several times larger for bright sources than for faint ones.
A SHAP analysis over the whole magnitude range would therefore mostly show the dependence on the flux and hide the smaller contributions of the other features.
To study the latter, we restrict the analysis to a narrow slice of $19.5 < r < 20.5$ in the object PSF magnitude, using $r$-band sources of point-like objects (extendedness of zero) only.

\begin{figure}
    \centering
    \includegraphics[width=0.6\textwidth]{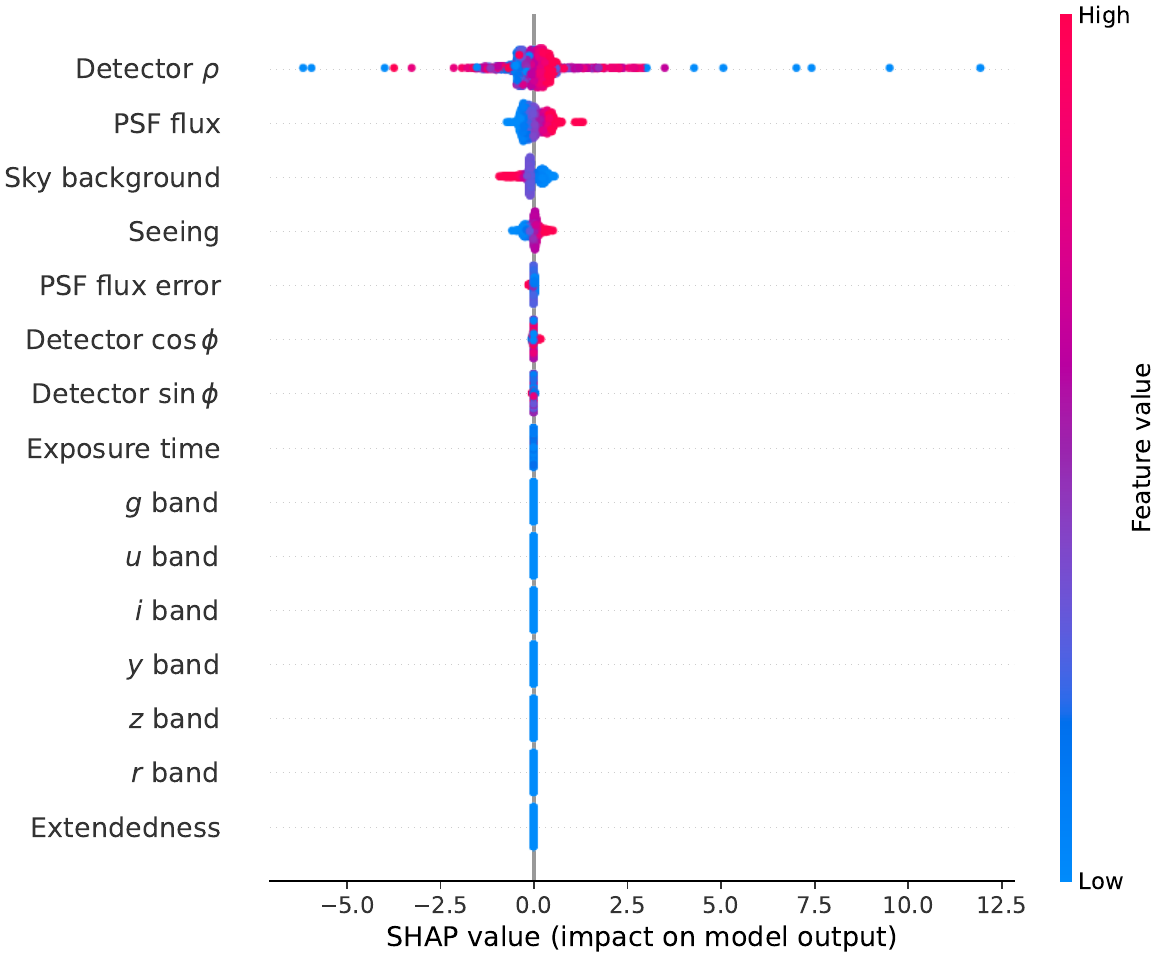}
    \caption{SHAP values of the input features for the uncertainty underestimation factor $u$ predicted by the MLP for the DP2 direct-image forced source photometry, for $r$-band sources of point-like objects with $19.5 < r < 20.5$ from the test set. Each dot is a source, positioned by the SHAP value of the given feature and colored by the feature value, from low (blue) to high (red). The color scale is linear and set separately for each feature, spanning the 5th to 95th percentiles of its values among the explained sources; values outside this range take the end colors, and the features that are constant within the sample, the passband and the extendedness, have a single color. Features are sorted by the mean absolute SHAP value.\label{fig:dp2-direct-shap}}
\end{figure}

Figure~\ref{fig:dp2-direct-shap} shows the SHAP values for the DP2 direct-image forced source photometry.
Within the magnitude slice, the detector position $\rho$ is the most important feature: for most sources it changes $u$ by less than 2.5, but for a few of them it adds up to 12 to $u$, which may indicate that the uncertainties of some detectors are underestimated much more than those of the others.
The source flux remains the second most important feature even within the one-magnitude slice, with larger $u$ for brighter sources, consistent with Section~\ref{sec:results-dp2}.
Higher sky background decreases $u$, as expected when the background noise, which is included in the reported uncertainties, dominates over the unaccounted errors, while worse seeing slightly increases $u$.
The other features have small contributions.

\subsection{Alternative loss function setups}\label{sec:discussion-losses}

\begin{figure*}
    \centering
    \setlength{\tabcolsep}{1pt}
    \begin{tabular}{ccccc}
        \footnotesize Epps--Pulley, accumulated &
        \footnotesize $\chi^2$, accumulated &
        \footnotesize $\chi^2$, averaged &
        \footnotesize KL, accumulated &
        \footnotesize KL, averaged \\
        \footnotesize\shortstack{$\average{\hat\chi^2} = 1.01$\\ $\average{\mathrm{KL}} = 0.070$} &
        \footnotesize\shortstack{$\average{\hat\chi^2} = 0.95$\\ $\average{\mathrm{KL}} = 0.084$} &
        \footnotesize\shortstack{$\average{\hat\chi^2} = 1.27$\\ $\average{\mathrm{KL}} = 0.092$} &
        \footnotesize\shortstack{$\average{\hat\chi^2} = 2.28$\\ $\average{\mathrm{KL}} = 0.340$} &
        \footnotesize\shortstack{$\average{\hat\chi^2} = 2.15$\\ $\average{\mathrm{KL}} = 0.294$} \\
        \includegraphics[width=0.19\textwidth, trim=479 2.5 193.6 187.5, clip]{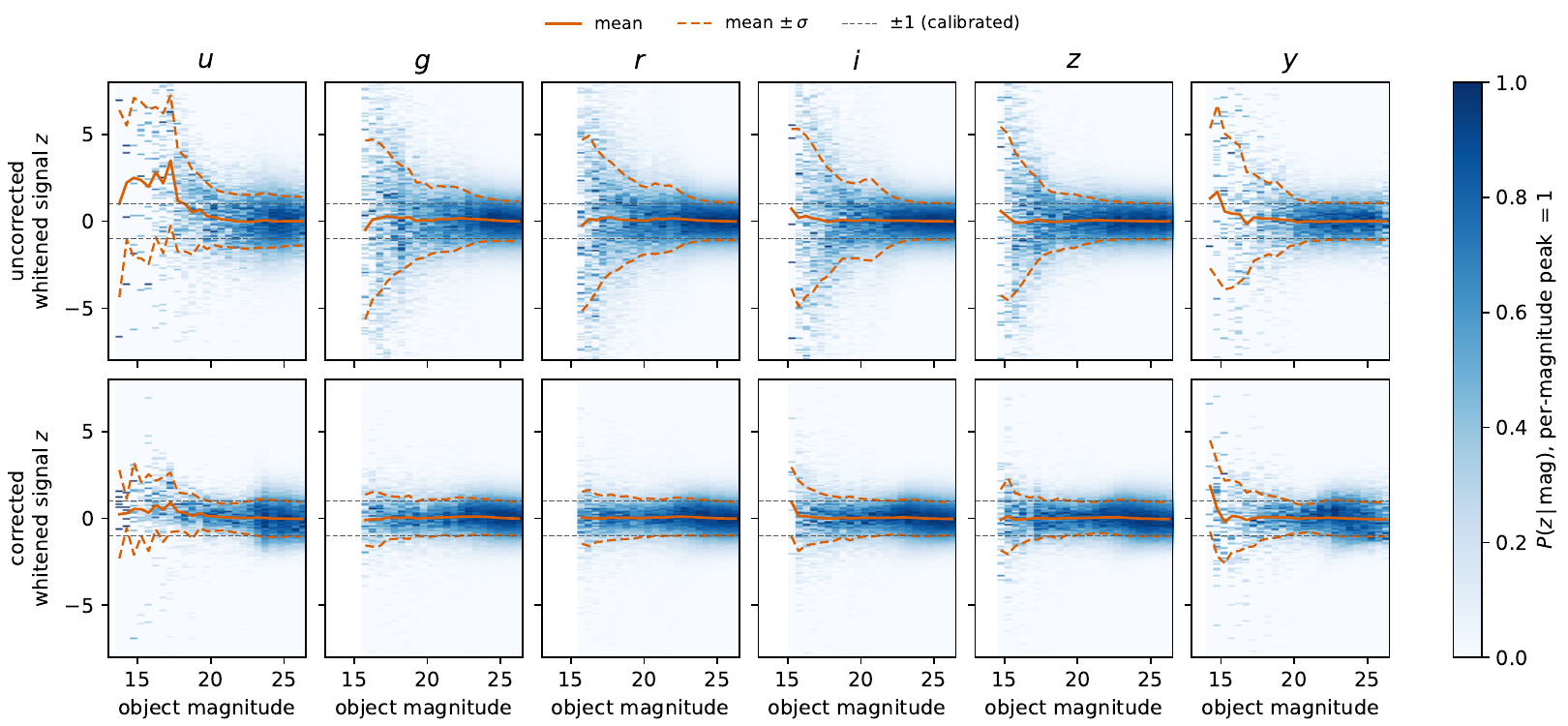} &
        \includegraphics[width=0.19\textwidth, trim=479 2.5 193.6 187.5, clip]{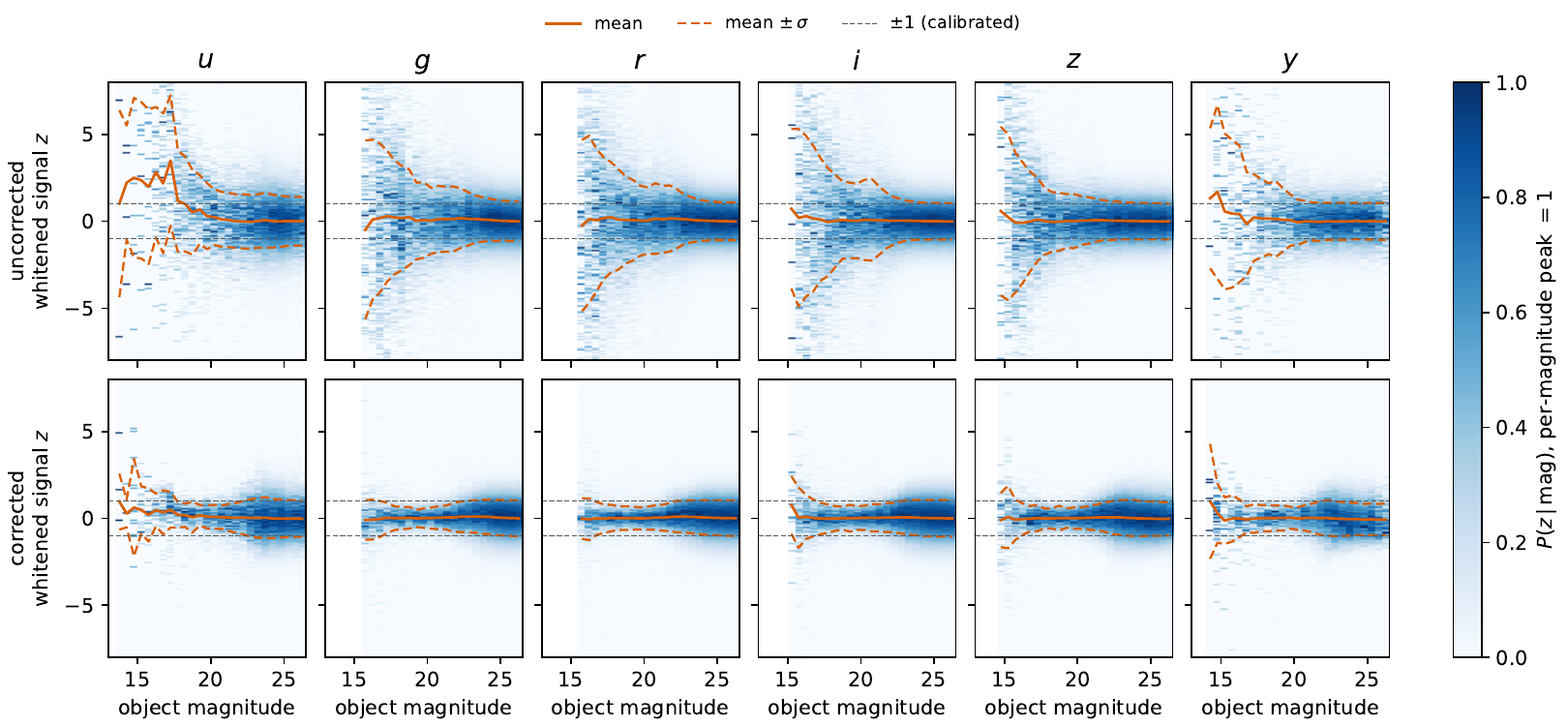} &
        \includegraphics[width=0.19\textwidth, trim=479 2.5 193.6 187.5, clip]{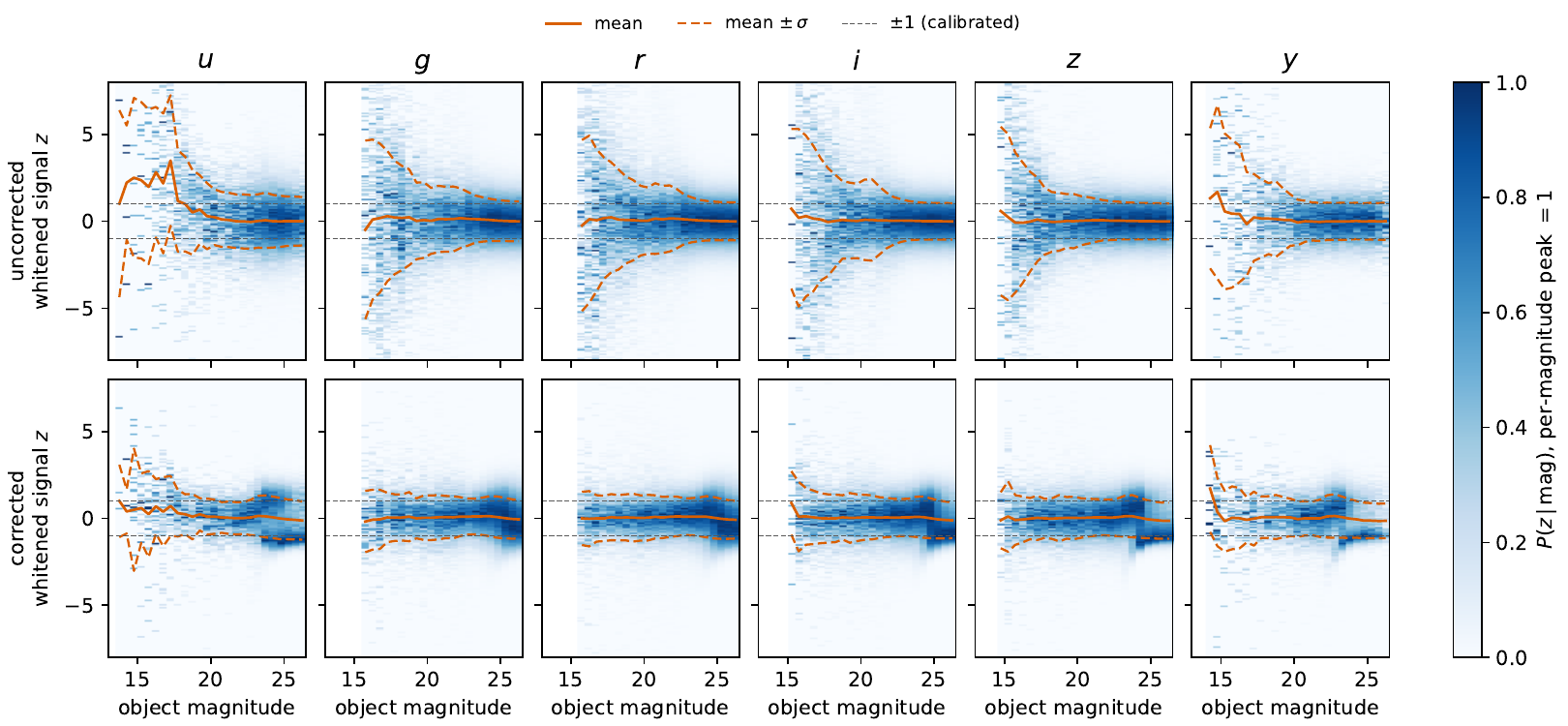} &
        \includegraphics[width=0.19\textwidth, trim=479 2.5 193.6 187.5, clip]{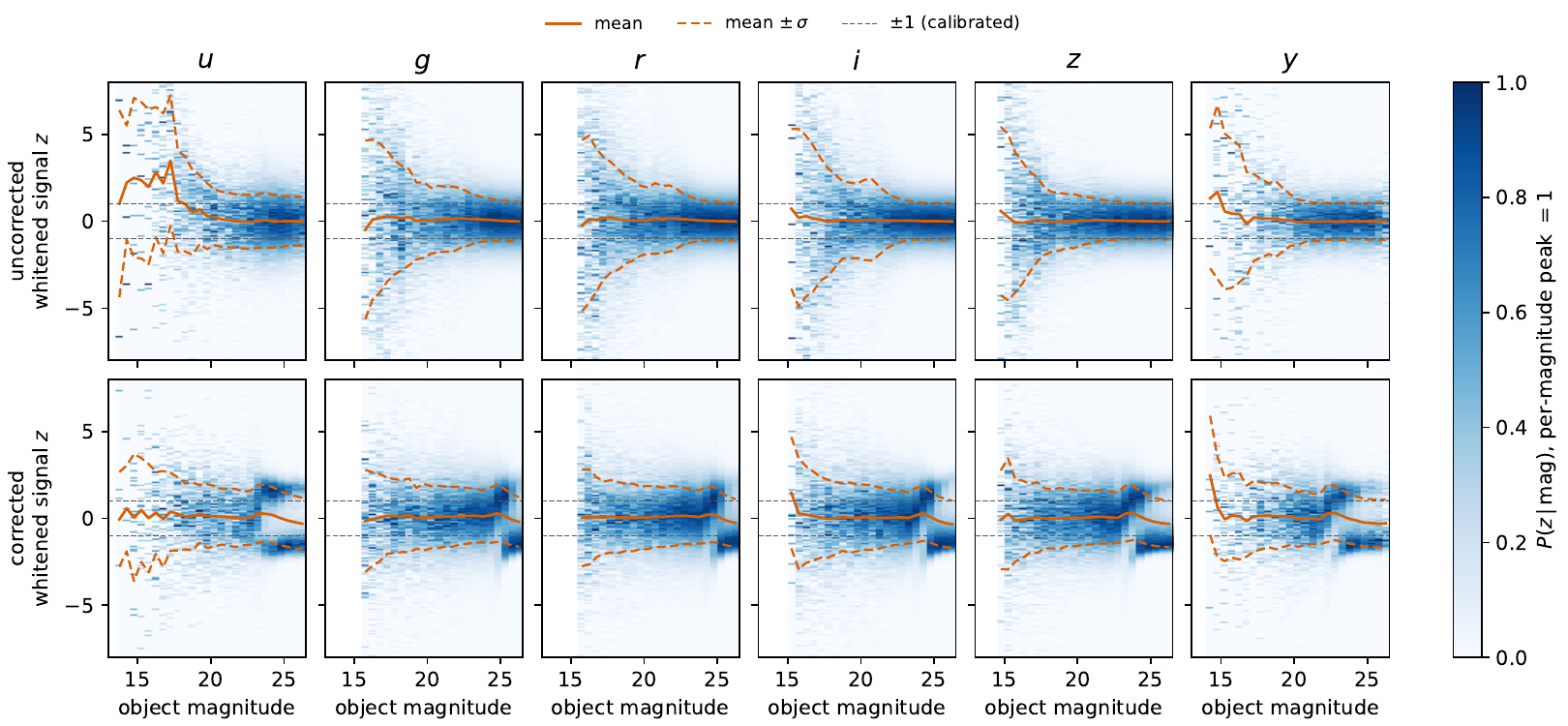} &
        \includegraphics[width=0.19\textwidth, trim=479 2.5 193.6 187.5, clip]{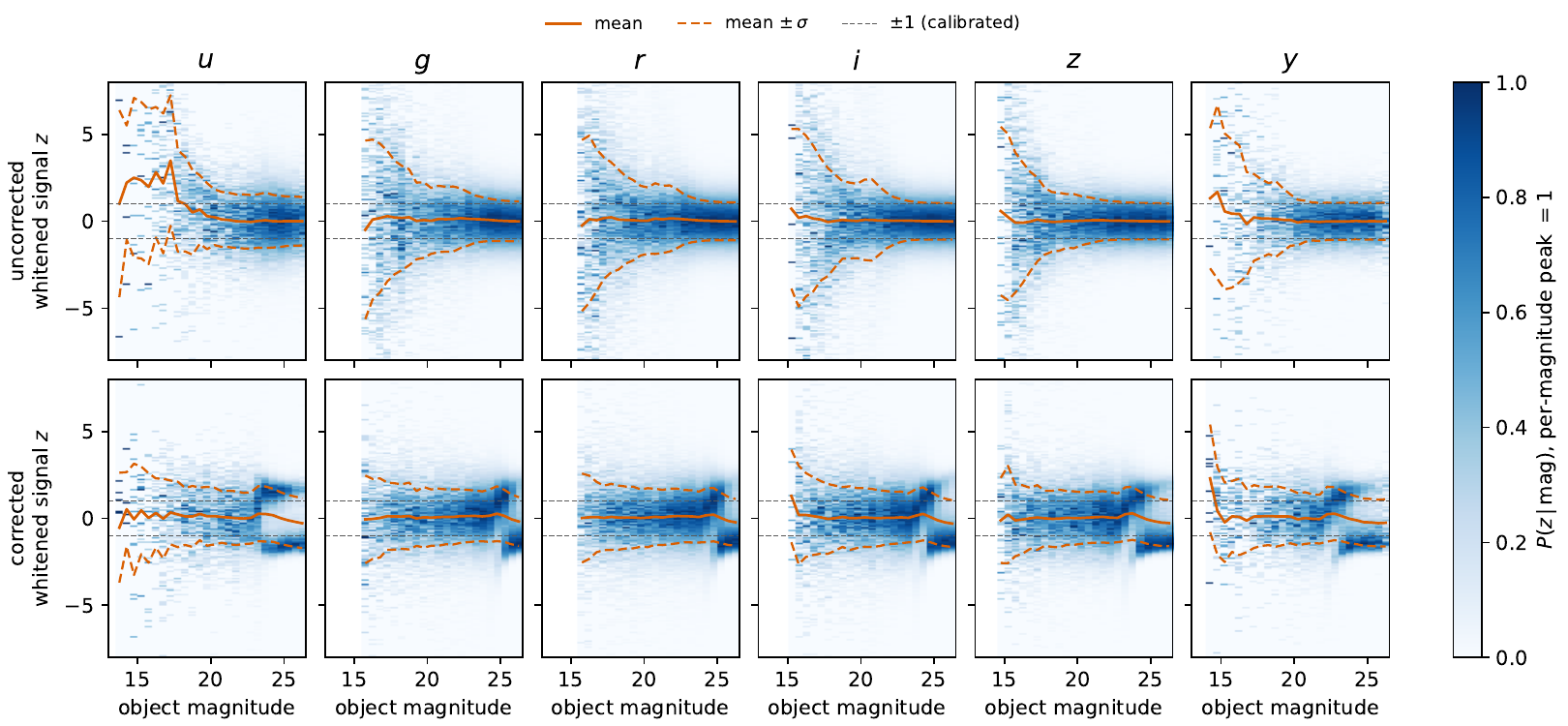} \\
    \end{tabular}
    \caption{Density of the corrected whitened signal as a function of the object PSF magnitude for the DP1 direct-image forced source photometry (test set) in the $z$-band, for the MLP models trained with different loss functions (columns), either accumulated over the batch or averaged over the light curves; the Epps--Pulley accumulated model is the fiducial one. The numbers above the panels are the mean reduced $\chi^2$ and the mean KL divergence of the light curves (Section~\ref{sec:results-dp2}), which are 1.24 and 0.126 with the reported uncertainties. The panels are cropped from the plots in the style of Figure~\ref{fig:dp2-direct-z} and share its color scale and vertical axis range: every magnitude column is normalized to its own peak, the solid curve is the mean, the dashed curves are the mean $\pm$ the standard deviation, and the thin dashed lines mark $\pm 1$.\label{fig:dp1-loss-z-mag}}
\end{figure*}

Each loss function of Section~\ref{sec:losses} may also be computed for each light curve separately and then averaged over the batch, instead of accumulating the statistic over all the whitened values of the batch.
To compare these setups, we train the DP1 direct-image MLP with the $\chi^2$ and KL losses, both accumulated and averaged, in the same way as the fiducial model, which uses the accumulated Epps--Pulley loss.
The results are generally consistent over the bands, so we show only the $z$-band, where the differences between the setups are the clearest.
Figure~\ref{fig:dp1-loss-z-mag} shows the mean reduced $\chi^2$ and the mean KL divergence of the light curves for each setup, together with the whitened signal as a function of magnitude.
Both $\chi^2$ setups improve the uncertainties compared to the reported ones, but less than the fiducial setup, while both KL setups make them even worse than the reported ones.
As a function of magnitude, the $\chi^2$ setups look not much worse than the fiducial one: the whitened signal has a nearly zero mean and a nearly unit standard deviation at most magnitudes, although the accumulated $\chi^2$ overestimates the uncertainties of bright sources.
The KL setups underestimate the uncertainties at all magnitudes, and the KL and the averaged $\chi^2$ setups show a sharp transition of the whitened signal distribution at about 23.5\,mag.

\begin{figure*}
    \centering
    \includegraphics[width=\textwidth]{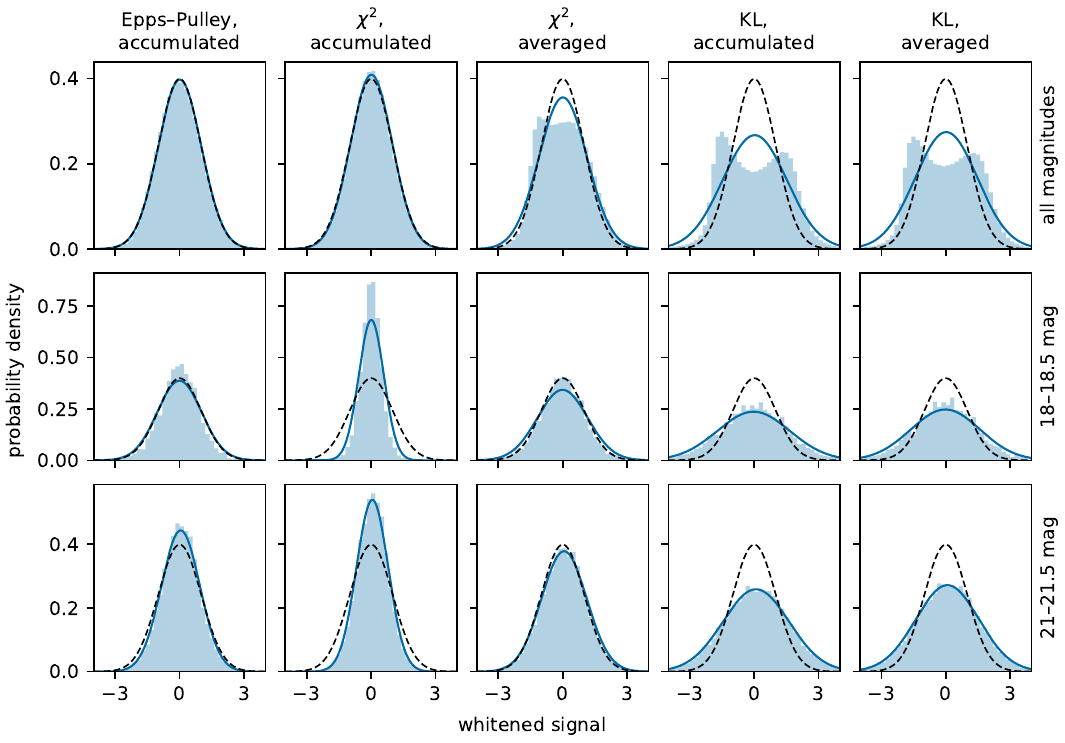}
    \caption{Distributions of the corrected whitened signal for the same models and band as in Figure~\ref{fig:dp1-loss-z-mag}, over all object magnitudes (top row) and in the 18--18.5\,mag (middle row) and 21--21.5\,mag (bottom row) object magnitude bins. Each panel shows the fitted normal distribution (solid) and the standard normal one (dashed).\label{fig:dp1-loss-hists}}
\end{figure*}

The distributions of the whitened signal reveal larger differences (Figure~\ref{fig:dp1-loss-hists}).
Both KL setups give bimodal distributions: the KL loss compares only the first two moments of the whitened signal distribution with those of the standard normal one, so it does not constrain the shape of the distribution, which may even become bimodal.
The averaged $\chi^2$ setup gives a flattened core, so its distribution is not normal either.
The accumulated $\chi^2$ setup is the closest competitor of the fiducial one: over all magnitudes, both give distributions close to the standard normal one.
However, the overall distribution is dominated by the numerous faint sources, for which the reported uncertainties are almost correct, so we also show the distributions in two brighter magnitude bins, where the correction is substantial.
In these bins, the accumulated $\chi^2$ setup gives significantly narrower distributions than the standard normal one, i.e. it overestimates the uncertainties of bright sources, while the fiducial setup stays much closer to the standard normal distribution.
The $\chi^2$ loss constrains only the mean square of all the whitened values of a batch, so the errors of the correction at different magnitudes may compensate each other, while the Epps--Pulley loss constrains the whole shape of the distribution, which disfavors such compensation.

\subsection{Alternatives to whitening}\label{sec:discussion-whitening}

We compare the whitening with three alternatives, which, as the whitened signal, have the standard normal distribution for each value under the null hypothesis.
All of them give the correct expectation of any statistic of the transformed values, so we compare their accuracy using the mean square of the transformed values.
The relative efficiency of a transformation is the ratio of the variances of the mean square for the whitening and for this transformation, i.e. approximately the ratio of the numbers of measurements they need to reach the same accuracy (Appendix~\ref{sec:alternatives-variance}); below, we quote the efficiencies for the DP2 uncertainties and $N = 100$ from Table~\ref{tab:efficiency}.
\begin{itemize}
    \item \emph{Pair differences} compare the fluxes of the object measured in two different visits, normalized by their combined uncertainty, so each visit is used in a single pair only, and they need approximately twice as many measurements as the whitening to reach the same accuracy, with an efficiency of 0.51.
    \item \emph{Hold-out mean} uses a part of the visits as a reference, similarly to a reference epoch, to estimate the mean flux of the object, and compares the fluxes measured in the other visits with it, taking the uncertainty of the reference flux into account.
    It is easy to implement and fast to compute; its accuracy depends on the hold-out fraction $q$, but even with the optimal choice, its efficiency is 0.69.
    \item \emph{Standardized residuals} are the relative residuals $r_{ki}$ (Eq.~\ref{eq:rel-res-sums}) normalized by their standard deviation, so that each of them is standard normal.
    The standardized residuals are as fast to compute as the whitened signal and are almost as accurate, unless the uncertainties within a light curve are so different that a few measurements dominate the weighted mean.
\end{itemize}
However, the standardized residuals are correlated within a light curve, while the whitened signal is independent, and the mean square of the whitened signal is the minimum-variance unbiased estimator of $u^2$ \citep[][Section~3.4]{lehmann1998theory}, because it is a function of the complete sufficient statistic.
Furthermore, when a single measurement dominates the weighted mean, the gradients of the standardized residuals with respect to the uncertainties become very large, while the gradients of the whitened signal stay bounded (Appendix~\ref{sec:light-curve-whitening}).
Therefore, we consider the whitening to be the preferable choice for both the statistical analysis and the gradient-based optimization.

\section{Conclusions}\label{sec:conclusions}

We have developed a method for validating and correcting photometric uncertainties with repeated measurements of non-variable objects.
Each light curve is transformed into a whitened signal, which follows the standard normal distribution for correct uncertainties regardless of the light curve length and of the spread of the uncertainties within it, and an uncertainty correction model is trained by matching the distribution of the whitened signal to the standard normal distribution with the Epps--Pulley statistic.
Among the transformations we considered, the whitening gives the most accurate estimate of the uncertainty scale: simple alternatives, such as pair differences or a reference subset of visits, need up to about twice as many measurements to reach the same accuracy, and the standardized residuals, which are almost as accurate, may have very large gradients, while the gradients of the whitened signal are always bounded.
Since the correction depends only on the properties of an individual measurement, it can also be applied to the measurements of variable objects, which cannot be used for the validation themselves.

As a demonstration, we applied the method to the Rubin DP2 forced source photometry.
For the direct-image photometry, the reported uncertainties are underestimated mostly for bright sources: by a factor of up to about 5 for objects brighter than 17--18\,mag, which corresponds to an additional magnitude error of about 10\,mmag.
The difference-image photometry is close to expected values, and its excess scatter is described by a constant magnitude error with no evidence of a color dependence, which goes down to 6--7\,mmag in Rubin-SV-225, the deepest of the fields we studied, indicating an excellent level of the photometric calibration.
At a fixed magnitude, the correction depends most strongly on the detector, which points to detector-dependent error terms as a direction for further investigation.

The method uses only the survey's own light curves of non-variable objects, without any external reference data, so it can be applied to other time-domain surveys as well.
Errors correlated between different light curves, such as errors of the photometric calibration of a visit, are beyond the scope of this work and are a natural extension of it.

\begin{acknowledgments}
We thank Jim~Bosch, Robert~Lupton, Tae-hyeon~Shin, and Michael~Wood-Vasey for valuable discussions.

This project is supported by Schmidt Sciences.
This material is based upon work supported in part by the National Science Foundation through Cooperative Agreements AST-1258333 and AST-2241526 and Cooperative Support Agreements AST-1202910 and 2211468 managed by the Association of Universities for Research in Astronomy (AURA), and the Department of Energy under Contract No. DE-AC02-76SF00515 with the SLAC National Accelerator Laboratory managed by Stanford University. Additional Rubin Observatory funding comes from private donations, grants to universities, and in-kind support from LSST-DA Institutional Members.
This publication is based in part on proprietary Rubin Observatory Legacy Survey of Space and Time (LSST) data, and was prepared in accordance with the Rubin Observatory data rights and access policies.
This work used the resources of the Rubin US Data Facility hosted by the SLAC Shared Science Data Facility (S3DF) at SLAC National Accelerator Laboratory, funded by Department of Energy under Contract No. DE-AC02-76SF00515.
This work used Bridges-2 \citep{Bridges2} at Pittsburgh Supercomputing Center through allocation PHY210095 from the Advanced Cyberinfrastructure Coordination Ecosystem: Services \& Support (ACCESS) program \citep{ACCESS}, which is supported by National Science Foundation grants \#2138259, \#2138286, \#2138307, \#2137603, and \#2138296.
In parts of this work, we used Claude Code with the Claude Opus and Claude Sonnet models, versions 4.7 to 5.5, for text generation, language editing, literature search, plotting of figures from existing data, and software development.
We also used GitHub Copilot for software development.
\end{acknowledgments}

\facilities{Rubin:Simonyi (LSSTComCam, LSSTCam), Rubin:USDAC}

\software{
\texttt{Dask} \citep{dask2016},
\texttt{HATS} \citep{2025arXiv250102103C},
\texttt{LSDB} \citep{2025arXiv250102103C},
\texttt{LSST Science Pipelines} \citep{2019ASPC..523..521B, 2026arXiv261022613O}, including the Data Butler \citep{2022SPIE12189E..11J},
\texttt{Matplotlib} \citep{2007CSE.....9...90H},
\texttt{nested-pandas} (\url{https://github.com/lincc-frameworks/nested-pandas}),
\texttt{Numba} \citep{2015llvm.confE...1L},
\texttt{NumPy} \citep{2020Natur.585..357H},
\texttt{pandas} \citep{2010scpy.soft.....M},
\texttt{PyArrow} \citep{apache_arrow},
\texttt{python-project-template} \citep{2024RNAAS...8..141O},
\texttt{PyTorch} \citep{paszke2019pytorch},
\texttt{SciPy} \citep{2020NaMet..17..261V},
\texttt{SHAP} \citep{lundberg2017shap},
\texttt{uncle-val} \citep{malanchev2026uncleval}.
}

\bibliography{uncle-val}{}
\bibliographystyle{aasjournalv7}

% Continue the main-text equation count in the appendix (A16, ...): with TeX Live >= 2025
% the AASTeX reset gives appendix equations the same hyperref targets as main-text ones
\edef\mainequations{\arabic{equation}}
\appendix
\setcounter{equation}{\mainequations}
\section{Light curve whitening}\label{sec:light-curve-whitening}

In this section, we derive the whitening operator, which transforms the vector $\bm{f}$ to $\bm{z}$, where $\bm{f}~\in~\mathbb{R}^N$, $\bm{f}~\sim~N(\mu, \diag{\sigma_1^2 \dots \sigma_N^2})$, $\bm{z}~\in~\mathbb{R}^{N-1}$, $\bm{z}~\sim~N(\bm{0}, I_{N-1})$.
Note that while we use the same notation in this appendix as in the main text, we omit the index $k$, because we are discussing a single light curve only, and use bold font to denote vectors: $\bm{x} \equiv (x_1 \dots x_N)^\T$.
We also assume that $\bm{\sigma}$ is the known vector of uncertainties, and $\mu$ is the unknown value of the ``true'' constant physical flux of the object, whose unbiased estimator is given by averaging the vector $\bm{f}$:
\begin{align}
\average{f} &\equiv \frac{\bm{w}^\T \bm{f}}{\bm{1}^\T \bm{w}}\,, \\
\bm{w} &\equiv \begin{pmatrix}\sigma_1^{-2} \\ \dots \\ \sigma_N^{-2}\end{pmatrix}\,,
\end{align}
see also Eq.~\eqref{eq:average-sums} for the definition via sums.

Since $\expect{\bm{f} - \average{f}} = \bm{0}$, the expectation of any linear transformation of the relative residuals $\bm{r}$ (Eq.~\ref{eq:rel-res-sums}) is zero: $\expect{A \bm{r}} = \bm{0}$ for any matrix~$A$.
The rest of this section is devoted to determining the transformation matrix $A$ that transforms $\bm{r}$ to $\bm{z}$.
Since the same whitening operator $A$ would transform the covariance matrix of the residual vector $\Sigma_r$ to the identity, i.e. $A \Sigma_r A^\T = I_{N-1}$, we will derive that covariance matrix first.

\subsection{Covariance matrix of the residual vector}\label{sec:residuals-covariance-matrix}
First, we derive the covariance matrix of the non-relative residuals $\bm{f} - \average{f}$:
\begin{equation}
    \Sigma_{\bm{f} - \average{f}} = \left(I_N - \frac{\bm{1} \bm{w}^\T}{\bm{1}^\T \bm{w}} \right) \diag{\sigma_1^2,\dots,\sigma_N^2} \left(I_N - \frac{\bm{1}\bm{w}^\T}{\bm{1}^\T \bm{w}} \right)^\T\,,
\end{equation}
where $\bm{1}$ is the all-ones column vector.
Skipping intermediate calculations:
\begin{equation}
    \Sigma_{\bm{f} - \average{f}} = \diag{\sigma_1^2,\dots,\sigma_N^2} - \frac1{\bm{1}^\T \bm{w}}J_N\,,
\end{equation}
where $J = \bm{1}\bm{1}^\T$ is the all-ones matrix.

Then the relative residuals are defined as $\bm{r} \equiv \diag{\sigma_1^{-1},\dots,\sigma_N^{-1}} (\bm{f} - \average{f})$ (see Eq.~\eqref{eq:rel-res-sums} for another form), so their covariance matrix is:
\begin{align}\label{eq:sigma_r}
    \Sigma_r &= \diag{\sigma_1^{-1},\dots,\sigma_N^{-1}} \, \Sigma_{\bm{f} - \average{f}} \, \diag{\sigma_1^{-1},\dots,\sigma_N^{-1}} = I_N - \bm{v}\bm{v}^\T\,, \\
    \bm{v'} &\equiv \begin{pmatrix}\sigma_1^{-1} \\ \dots \\ \sigma_N^{-1}\end{pmatrix}\,, \\
    \bm{v} &\equiv \frac{\bm{v'}}{\|\bm{v'}\|}\,,
\end{align}
where $\bm{v}$ is the unit vector proportional to the reciprocals of the uncertainties.
This result is well known for linear least squares: $\Sigma_r$ is often referred to as the residual-maker, or hat, matrix, which projects onto the hyperplane orthogonal to $\bm{v}$ \citep[see, e.g.,][]{hoaglin1978hat}.

\subsection{Whitening operator}
We are looking for an operator $A$ that transforms $\bm{r}$ to $\bm{z}$.
The matrix representing this operator must satisfy the following requirements:
\begin{itemize}
    \item $A$ is an $(N-1) \times N$ matrix, because $\bm{r}$ has length $N$, while the whitened vector $\bm{z}$ has one fewer component due to the unknown parameter $\mu$.
    \item $\bm{z}$ is standard multivariate normal, so $A \Sigma_r A^\T = \Sigma_z = I_{N-1}$.
    \item It is convenient for $A$ to be row-orthonormal: $A A^\T = I_{N-1}$.
    \item The last two requirements also imply $\Sigma_r = A^\T A$, i.e. $A$ is a square-root decomposition of $\Sigma_r$, which exists since covariance matrices are positive semidefinite.
\end{itemize}

We rewrite the decomposition using Eq.~\eqref{eq:sigma_r}:
\begin{align}\label{eq:householder}
    I_N &= \bm{v}\bm{v}^\T + \Sigma_r = \bm{v}\bm{v}^\T + A^\T A = \left(-\bm{v}, {A}\right) \begin{pmatrix}-\bm{v}^\T \\ A\end{pmatrix} = H^\T H\,, \\
    H &\equiv \begin{pmatrix}-\bm{v}^\T \\ A\end{pmatrix}\,.
\end{align}
This means that if we find a matrix $H$ that is orthogonal and has the desired form, we can simply drop its first row to obtain the whitening operator.
The Householder reflection matrix satisfies all these criteria if the Householder vector is chosen as $\bm{h} = \bm{v} + \bm{e}_1$, where $\bm{e}_1$ is the first unit coordinate vector~\citep{householder:hal-01316095, GolubVanLoan2013}:
\begin{equation}
    H = I_N - 2 \frac{\bm{h}\bm{h}^\T}{\bm{h}^\T\bm{h}}\,.
\end{equation}
This choice of $\bm{h}$ can be described geometrically: since the Householder operator reflects any vector across the plane perpendicular to $\bm{h}$, the reflection of $\bm{e}_1$ is $-\bm{v}$, i.e. $H \bm{e}_1 = -\bm{v}$; this is also the reason we used the minus sign in Eq.~\eqref{eq:householder}.

Finally, the flux whitening operator ($\bm{z} = M \bm{f}$) is:
\begin{equation}\label{eq:whitening-operator}
    M = \left(I_N - 2 \frac{\bm{h}\bm{h}^\T}{\bm{h}^\T\bm{h}}\right)_{2:N,:} \, 
    \diag{\sigma_1^{-1},\dots,\sigma_N^{-1}} \,
    \left(I_N - \frac{\bm{1}\bm{w}^\T}{\bm{1}^\T\bm{w}}\right) \,.
\end{equation}
Note that the only term depending on the scale of the uncertainties is the diagonal matrix.
That makes this operator linear in the constant multiplicative factor applied to $\bm{\sigma}$, which means that a multiplicative change of the uncertainties $\bm{\sigma} \rightarrow u \cdot \bm{\sigma}$ would produce a multiplicative change of the whitened signal $\bm{z} \rightarrow 1/u \cdot \bm{z}$.
Also, $\bm{z}^\T \bm{z} = \bm{r}^\T A^\T A \bm{r} = \bm{r}^\T \Sigma_r \bm{r} = \bm{r}^\T \bm{r} \sim \chi^2_{N-1}$, as $\bm{z}$ consists of $N - 1$ independent standard normal values.

It is easy to see that the proposed matrix $M$ is not the only one satisfying the criteria listed above: the choice of the unit coordinate onto which $-\bm{v}$ is reflected was arbitrary.
As usual in linear algebra, if there is more than one solution, then there are infinitely many; for example, we can do the same job with the eigenvalue decomposition, dropping the row in the decomposition matrix corresponding to the zero eigenvalue.
However, the eigendecomposition has $\Theta(N^3)$ computational complexity, and the gradient calculation becomes unstable as it approaches all-but-one unit eigenvalues.
In contrast, the proposed transformation does not require constructing the matrix explicitly: the whitened signal consists of the last $N - 1$ components of $H \bm{r} = \bm{r} - 2 \bm{h} (\bm{h}^\T \bm{r}) / (\bm{h}^\T \bm{h})$, so it has $\Theta(N)$ complexity.
It also leads to stable gradient calculations, because the only denominator, $\bm{h}^\T \bm{h} = 2 (1 + v_1)$, is not smaller than $2$ for any uncertainties.

Practically, the choice of the unit vector onto which $-\bm{v}$ is reflected may cause a bias in a statistical analysis, due to numerical artifacts, so we recommend either adapting the algorithm to select a random axis or pre-shuffling the data.
In our analysis we technically do the former by randomly sampling a fixed number of sources per light curve.

\section{Statistical efficiency of light curve transformations}\label{sec:alternatives-variance}

In this section, we compare the whitening with alternative transformations of a light curve, which also produce values with the standard normal marginal distribution under the null hypothesis.
We use the same notation as in Appendix~\ref{sec:light-curve-whitening}, omitting the light curve index $k$.
Let a transformation produce a vector $\bm{t}~\in~\mathbb{R}^n$, $\bm{t}~\sim~N(\bm{0}, C)$, where $C$ is the correlation matrix, i.e. $C_{ii} = 1$.
The simplest statistic is the mean square
\begin{equation}\label{eq:mean-square}
    S \equiv \frac1n \sum_{i=1}^n t_i^2\,,
\end{equation}
which is an estimator of $u^2$; under the null hypothesis $\expect{S} = 1$ for any transformation of this kind, so the transformations differ only in the variance of $S$.
We characterize each transformation by its relative efficiency with respect to the whitening,
\begin{equation}\label{eq:relative-efficiency}
    e \equiv \frac{\variance{S_\mathrm{whiten}}}{\variance{S}}\,,
\end{equation}
which for large $N$ is the ratio of the numbers of measurements required by the whitening and by the given transformation to reach the same accuracy.
To find its variance, we write
\begin{equation}
    \variance{\bm{t}^\T \bm{t}} = \expect{\left(\sum_{i=1}^n t_i^2\right)^2} - \left(\expect{\sum_{i=1}^n t_i^2}\right)^2 = \sum_{i,l} \expect{t_i^2 t_l^2} - n^2\,,
\end{equation}
which requires the fourth moments of the jointly normal values.
For jointly normal values, $\expect{t_i^2 t_l^2} = C_{ii} C_{ll} + 2 C_{il}^2 = 1 + 2 C_{il}^2$ \citep{1899RSPTA.192..101S}, a special case of Isserlis' theorem \citep{isserlis1918formula}.
Summing over $i$ and $l$ and using the symmetry of $C$, we obtain $\variance{\bm{t}^\T \bm{t}} = 2 \sum_{i,l} C_{il}^2 = 2 \operatorname{tr}(C^2)$, and
\begin{equation}\label{eq:mean-square-variance}
    \variance{S} = \frac{2 \operatorname{tr}(C^2)}{n^2} = \frac2n \left(1 + \frac1n \sum_{i \neq l} C_{il}^2 \right)\,,
\end{equation}
i.e. any correlation between the components of $\bm{t}$ increases the variance.

\subsection{Whitening}
The whitened signal has $C = I_{N-1}$ and $n = N - 1$, so
\begin{equation}
    \variance{S_\mathrm{whiten}} = \frac2{N-1}
\end{equation}
for any $\bm{\sigma}$.

\subsection{Pair differences}
If we split the light curve into $\lfloor N/2 \rfloor$ disjoint pairs $(i, l)$, the values
\begin{equation}
    t_{il} = \frac{f_i - f_l}{\sqrt{\sigma_i^2 + \sigma_l^2}}
\end{equation}
do not depend on $\mu$ and are independent, so $C = I$, $n = \lfloor N/2 \rfloor$, and
\begin{equation}
    \variance{S_\mathrm{pairs}} = \frac2{\lfloor N/2 \rfloor}\,, \quad
    e_\mathrm{pairs} = \frac{\lfloor N/2 \rfloor}{N - 1} \approx \frac12\,.
\end{equation}

\subsection{Hold-out mean}
If we estimate $\mu$ from a subset $\mathcal{H}$ of $n_\mathcal{H}$ measurements and use the remaining subset $\bar{\mathcal{H}}$ of $n = N - n_\mathcal{H}$ measurements, the values are
\begin{equation}
    t_i = \frac{f_i - \average{f}_\mathcal{H}}{\sqrt{\sigma_i^2 + 1/W_\mathcal{H}}}\,, \quad i \in \bar{\mathcal{H}}\,, \quad W_\mathcal{H} \equiv \sum_{i \in \mathcal{H}} \sigma_i^{-2}\,,
\end{equation}
where $\average{f}_\mathcal{H}$ is the weighted mean over $\mathcal{H}$, which has the variance $1/W_\mathcal{H}$.
All the values share the same $\average{f}_\mathcal{H}$, so they are positively correlated:
\begin{equation}
    C_{il} = c_i c_l\,, \quad c_i^2 \equiv \frac1{1 + \sigma_i^2 W_\mathcal{H}}\,,
\end{equation}
and
\begin{equation}\label{eq:mean-square-variance-decomp}
    \variance{S_\mathrm{hold}} = \frac2{n^2} \left[n + \left(\sum_{i \in \bar{\mathcal{H}}} c_i^2\right)^2 - \sum_{i \in \bar{\mathcal{H}}} c_i^4\right]\,.
\end{equation}
For equal uncertainties, $c_i^2 = 1/(n_\mathcal{H} + 1)$, and $\variance{S_\mathrm{hold}} = \frac2n\left[1 + \frac{n - 1}{(n_\mathcal{H} + 1)^2}\right]$.
In a more general case, if $\sigma_i^2 W_\mathcal{H} \gg 1$ for all $i \in \bar{\mathcal{H}}$, the sum $\sum_{i \in \bar{\mathcal{H}}} c_i^2 \approx W_{\bar{\mathcal{H}}} / W_\mathcal{H}$, and for a hold-out fraction $q \equiv n_\mathcal{H}/N$:
\begin{equation}\label{eq:hold-out-efficiency}
    \frac1{e_\mathrm{hold}} \approx \frac1{1 - q} + \frac1{q^2 N}\,,
\end{equation}
where we used $W_{\bar{\mathcal{H}}} / W_\mathcal{H} \approx (1 - q)/q$.
For large $N$, the efficiency has a maximum at
\begin{equation}\label{eq:hold-out-optimal}
    q_\mathrm{opt} \approx \left(\frac2N\right)^{1/3}\,, \quad
    e_\mathrm{hold} \approx 1 - \frac32 \left(\frac2N\right)^{1/3}\,.
\end{equation}
Therefore, we recommend choosing the hold-out fraction close to $(2/N)^{1/3}$ rather than a fixed value.

\subsection{Standardized residuals}
From Eq.~\eqref{eq:sigma_r}, the variance of the relative residual is $\variance{r_i} = 1 - v_i^2$, so the standardized residuals are
\begin{equation}\label{eq:standardized-residuals}
    t_i = \frac{r_i}{\sqrt{1 - v_i^2}} = \frac{f_i - \average{f}}{\sqrt{\sigma_i^2 - 1/(\bm{1}^\T \bm{w})}}\,,
\end{equation}
with $n = N$.
The off-diagonal elements of the correlation matrix are
\begin{equation}
    C_{il} = -c_i c_l\,, \quad c_i^2 \equiv \frac{v_i^2}{1 - v_i^2}\,,
\end{equation}
and Eq.~\eqref{eq:mean-square-variance} gives
\begin{equation}\label{eq:std-variance}
    \variance{S_\mathrm{std}} = \frac2{N^2} \left[N + \left(\sum_{i=1}^N c_i^2\right)^2 - \sum_{i=1}^N c_i^4\right]\,.
\end{equation}

Since $\sum_{i=1}^N v_i^2 = 1$, for equal uncertainties $v_i^2 = 1/N$ and $c_i^2 = 1/(N-1)$, so $\variance{S_\mathrm{std}} = 2/(N-1) = \variance{S_\mathrm{whiten}}$.
In the general case, assuming that no single measurement dominates the weighted mean, $v_i^2 = O(1/N)$, and expanding to the leading order:
\begin{equation}\label{eq:std-efficiency}
    e_\mathrm{std} = 1 - \frac{\kappa - 1}{N^2} + O(N^{-3})\,, \quad \kappa \equiv N \sum_{i=1}^N v_i^4 = O(1)\,.
\end{equation}

\subsection{Examples of relative efficiencies}
Table~\ref{tab:efficiency} gives the relative efficiencies of all the transformations for three cases: equal uncertainties, exponentially distributed uncertainties, and the reported uncertainties of Rubin DP2.
For each case, we average the exact variances of $S$ (Eqs.~\eqref{eq:mean-square-variance-decomp} and~\eqref{eq:std-variance}) over the light curves, substitute them into Eq.~\eqref{eq:relative-efficiency}, and choose the hold-out fraction $q_\mathrm{opt}$ maximizing the resulting efficiency, with the hold-out subset selected randomly.

For equal uncertainties, the standardized residuals are as efficient as the whitening, and the efficiency of the hold-out mean is given by its exact expression for equal uncertainties.

The approximations in Eqs.~\eqref{eq:hold-out-efficiency}--\eqref{eq:hold-out-optimal} and~\eqref{eq:std-efficiency} assume that $\expect{\sigma^{-2}}$ exists, so that no single measurement dominates the weighted mean.
To demonstrate a case where it does not exist, we take $\sigma$ exponentially distributed, which is a strongly skewed distribution, and average over random realizations of $\bm{\sigma}$.

For the real uncertainties, we take the DP2 forced direct-image photometry of the objects brighter than 21\,mag in the given band from a subset of the catalog, with the same quality cuts and rejection of variable objects as for the training data (Sections~\ref{sec:data} and~\ref{sec:rejection-of-variable-objects}).
Every light curve with at least $N$ measurements is randomly subsampled to $N$ measurements, which gives 180\,783, 4\,147, and 651 light curves for $N = 10$, $30$, and $100$, respectively; DP2 has no single-band light curves with $N = 1000$ measurements.

\begin{deluxetable}{lcccc}
\tablecaption{Relative efficiencies $e$ of the light curve transformations with respect to the whitening for equal, exponentially distributed, and Rubin DP2 uncertainties.\label{tab:efficiency}}
\tablehead{\colhead{Transformation} & \colhead{$N = 10$} & \colhead{$N = 30$} & \colhead{$N = 100$} & \colhead{$N = 1000$}}
\startdata
\cutinhead{Equal uncertainties}
Whitening & $1$ & $1$ & $1$ & $1$ \\
Pair differences & $0.56$ & $0.52$ & $0.51$ & $0.50$ \\
Hold-out mean, $q = q_\mathrm{opt}$ & $0.57$ & $0.60$ & $0.69$ & $0.83$ \\
Standardized residuals & $1$ & $1$ & $1$ & $1$ \\
\cutinhead{Exponentially distributed uncertainties}
Whitening & $1$ & $1$ & $1$ & $1$ \\
Pair differences & $0.56$ & $0.52$ & $0.51$ & $0.50$ \\
Hold-out mean, $q = q_\mathrm{opt}$ & $0.52$ & $0.56$ & $0.65$ & $0.81$ \\
Standardized residuals & $0.95$ & $0.98$ & $0.994$ & $0.9994$ \\
\cutinhead{DP2 uncertainties}
Whitening & $1$ & $1$ & $1$ & \nodata \\
Pair differences & $0.56$ & $0.52$ & $0.51$ & \nodata \\
Hold-out mean, $q = q_\mathrm{opt}$ & $0.55$ & $0.60$ & $0.69$ & \nodata \\
Standardized residuals & $0.996$ & $0.9998$ & $0.99997$ & \nodata \\
\enddata
\tablecomments{The optimal hold-out fractions for $N = 10$, $30$, $100$, and $1000$ are $q_\mathrm{opt} = 0.30$, $0.30$, $0.22$, and $0.12$ for the equal uncertainties, $0.40$, $0.33$, $0.25$, and $0.13$ for the exponentially distributed uncertainties, and $0.30$, $0.30$, and $0.22$ for the DP2 uncertainties.}
\end{deluxetable}

\end{document}